\documentclass[ 12pt,draftclsnofoot,onecolumn]{IEEEtran}

\usepackage[T1]{fontenc} 
\usepackage{url}

\usepackage{color}
\usepackage{subcaption}
\usepackage{amsmath,amssymb,amsfonts}
\usepackage{algorithmic}
\usepackage{graphicx}
\usepackage{textcomp}
\usepackage{booktabs}
\usepackage{cite}
\usepackage{caption}
\usepackage{mathtools}
\usepackage{multicol}
\usepackage[latin1]{inputenc}
\usepackage{mathtools}

\usepackage{float}

\usepackage{amsmath}

\usepackage{fixltx2e}

\begin{document}
	\title{Viral Aerosol Concentration Characterization and Detection in Bounded Environments}
	\author{{Osama Amin, Hayssam Dahrouj, Nojood Almayouf, Tareq Y. Al-Naffouri, Basem Shihada  and Mohamed-Slim~Alouini}

		\thanks{ O. Amin, H. Dahrouj, N. Almayouf, T. Y. Al-Naffouri, B. Shihada and M.-S. Alouini are with CEMSE Division,  King Abdullah University of Science and Technology (KAUST),  Thuwal, Makkah Province, Saudi Arabia. E-mail: \{osama.amin, hayssam.dahrouj, nojood.almayouf@kaust.edu.sa, tareq.alnaffouri, basem.shihada, slim.alouini\}@kaust.edu.sa.}}
	
	\maketitle
	\begin{abstract}
		Viral spread has been intermittently threatening human life over time. Characterizing the viral concentration and modelling the viral transmission are, therefore, considered major milestones for enhancing viral detection capabilities. This paper addresses the problem of viral aerosol detection based on the exhaled breath in a bounded environment, e.g., a bounded room. The paper models the exhaled breath as a cloud which is emitted through the room continuously, and analyzes the temporal-spatial virus concentration by accounting for partial absorption and reflection at each side of the room. The paper first derives a closed form expression of the temporal-spatial virus concentration. It then considers the deployment of a receiver composed of an air sampler and a bio-sensor to detect the viral existence of a specific virus. We, therefore, assess the detection capabilities of the proposed system via evaluating the viral miss-detection probability as a function of the sampling volume and the detection time-instance at the receiver side. Our numerical simulations verify the validity of the analytical results, and illustrate the ability of the proposed system to detect viruses in indoor environments. The results further characterize the impacts of several system parameters on the miss-detection probability.
	\end{abstract}


	\section{Introduction}
	\IEEEPARstart{V}{iral} spread into the respiratory system often leads to life-threatening infections, the escalation of which could reach epidemic or pandemic levels. Airborne transmission of pathogens has recently received considerable attention across different research boards because of their epidemic impacts on human life. The world has indeed suffered from various viral pandemic waves over the past century, such as 1918-1920 influenza, 1957-1958 H2N2 virus, 1968 H3N2 virus, 2009 H1N1 \cite{online_pandemics}, and most recently the COVID-19 (also known as SARS-COV-2) \cite{rudan2020cascade}. The 1918 flu, which was first detected in the United States in March 1918, went through three different waves before fading out in Summer 1919. It was estimated that about one-third of the world population was infected with the 1918 flu, with at least 50 million fatalities \cite{online_pandemics}. H2N2, also known as the new influenza A virus, appeared in February 1957 in East Asia and reached the United States in Summer 1957, resulting in about 1.1 million deaths worldwide. H3N2 was first observed in the United States in September 1968. It then spread worldwide, leading to around 1 million deaths. The 2009 H1N1 pandemic has the first reported case on April 2009. It went afterwards through different waves until it disappeared in August 2010 \cite{online_pandemics}. Nowadays, the world is facing another severe viral pandemic, i.e., COVID-19, whose first reported cases were in China in December 2019 \cite{rudan2020cascade}. As of June 2020, the statistics show that COVID-19 already infected 10 million people worldwide, with around 500 thousand fatalities. Such numbers are expected to grow further in the near future, especially given the absence of vaccines and the projected potential waves of the virus. It is, therefore, necessary to steer parts of the scientific research towards combatting similar viral crises by means of properly reducing pandemic fierceness. This paper addresses one particular aspect of viral transmission in bounded environments. It characterizes the temporal-spatial virus concentration in an effort towards enhancing viral detection capabilities.
	

	The fundamental research on viral spread is related to several multi-disciplinary areas, which include treatment, vaccine development, diagnostics, epidemic propagation, finance and economy, safety and public health, and crisis management. Investigating spatial-temporal viral spread performance from respiration, coughing, and sneezing in a bounded environment is one particular relevant research direction which needs further understanding and investigation. The rationale behind such study is to help estimating the probability of infection in indoor environments so as to detect specific viruses across time and space. Such direction is also promising for mitigating the viral spread, and identifying possible early hazards. Characterizing spatial-temporal viral spread is also critical for investigating initiatives that reduce anticipated viral waves, which often follow the first wave. Such initiatives indeed aim at stopping any potential epicenter that would initiate subsequent waves. For example, mass gathering events can easily trigger an epicenter of an epidemic wave, such as the one that occurred during the Champions League soccer game that was held in Italy in February 2020, where is considered as COVID-19 ``biological bomb'' in Spain \cite{rudan2020cascade}. Detecting infected cases in mass gathering events can, therefore, thwart possible traces of viral threats. Developing powerful mathematical tools for understanding and analyzing virus transmission can then provide paramount guidelines to reduce infection rates and detect viruses, especially when equipped with the appropriate sensors. This paper focuses on one special aspect of communication via breath in bounded environments, and analyzes the temporal-spatial virus concentration of the underlying system.
	
	\subsection{Related Work}
	
	Communication via breath, recently introduced under the umbrella of molecular communication, describes scenarios where several data can be exchanged through the inhalation and exhalation processes \cite{khalid2019communication, amin2020airborne, khalid2018system, khalid2020modeling}. Studying viruses spread and detecting viruses from the exhaled breath form one aspect of this research area, where viruses transmitted through aerosol channel can be detected using suitable biological receivers (sensors). The investigation of such an interested problem from an engineering perspective push the molecular communication research community to adopt some of established techniques to better understand and analyze communication via breath-based systems \cite{khalid2018system, khalid2020modeling}. In \cite{khalid2018system}, the problem of detecting viruses from the exhaled breath is analyzed by assuming largely unbounded environment. In this context, the steady state analysis is considered and the probability of miss-detection is analyzed assuming that a Silicon Nano-wire field effect transistor is used for detection. In \cite{khalid2020modeling}, the study is extended to include multiple sources, such as sneezing and coughing, in addition to the introduction of transient analysis and frequency response. Both studies \cite{khalid2018system, khalid2020modeling} are suitable for large rooms where one can neglect the effect of boundary absorption and reflection. Furthermore, both studies model the exhaled breath as a point source that is subjected to airflow with a constant velocity. Although this model can give an acceptable approximation when the airflow velocity is high, it is not suitable for situations where there is no external air flow, and where the exhaled nasal or mouth breath pattern would follow specific patterns. In fact, in an experimental setup \cite{tang2013airflow}, an image-based study investigates the spatial-temporal pattern of human jet sources, which come from breathing, coughing and sneezing. In \cite{tang2013airflow}, the images of a high-speed camera show a relatively symmetrical, conical geometry and measure the expansion rate and area, which justifies parts of the system model and assumptions adopted in our paper.
	
	\subsection{Contributions}
	Unlike the aforementioned references, this paper considers a bounded-room where the ceil, floor, and walls can partially or fully reflect/absorb the emitted viral aerosol particles. We analyze the spatial-temporal diffused viral concentration in the room, assuming a single source of the exhaled breath. In this regard, we adopt an accurate model for the exhaled breath as a conical cloud released from the nose or the mouth in a room space. Then, we derive a closed-form expression for the virus concentration in time and space inside the room. Finally, we analyze the ability to detect the virus by taking spatial samples using a suitable bio-sensor. Towards this direction, the paper assesses the detection capabilities of the proposed system via evaluating the viral miss-detection probability as a function of the sampling volume and the detection time-instance at the receiver side. The paper numerical simulations verify the validity of the analytical results, and illustrate the ability of the proposed system to detect viruses in indoor environments. The results further characterize the impacts of the indoor system parameters (e.g., absorption and reflection coefficients, room dimensions, etc.) on the miss-detection probability.
	
	\subsection{Organization}
	The rest of this paper is organized as follows, Section II provides a description of the indoor system model. Section III analyzes the spatial-temporal viral concentration due to a point source. Then, Section IV derives the  spatial-temporal viral concentration due to an exhalation. Section V analyzes the probability of missed detection using an appropriate bio-sensor. Finally, Section VI assesses the system performance followed by the paper conclusion in Section VII.

	\section{System Description}
	
\begin{figure}[t]
		\centering
		\includegraphics[width=5.5in]{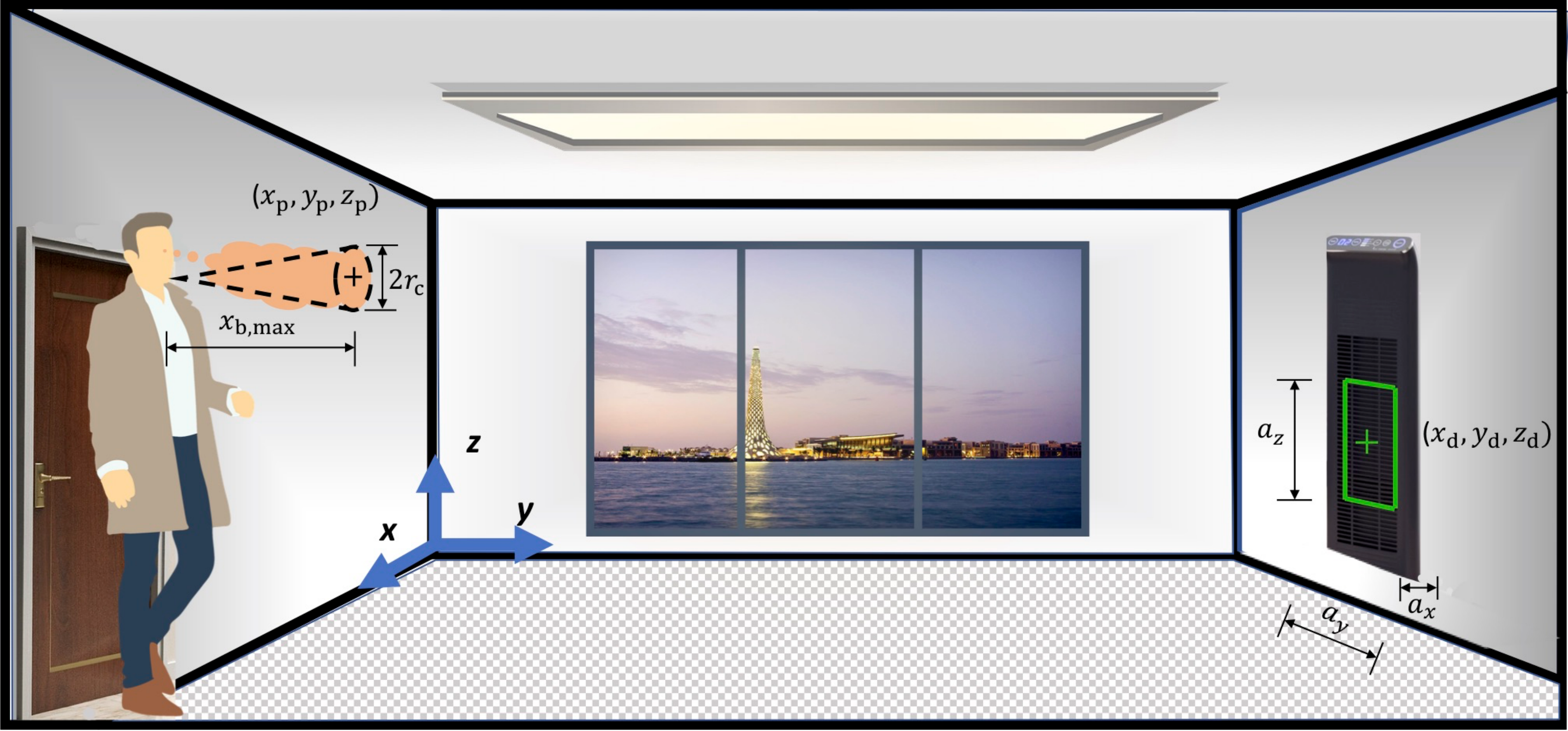}
		\caption{Bounded environment indoor model.} \label{fig:sys_model}
	\end{figure}

	Consider a 3-D bounded room in $x$, $y$ and $z$ dimensions, as depicted in Fig. \ref{fig:sys_model}, where $0 \leq x \leq L_x$, $0 \leq y \leq L_y$ and $0 \leq z \leq L_z$, where $L_x$, $L_y$, and $L_z$ are fixed positive constants. In the room, we assume the presence of a person infected with a specific virus, where virus particles spread in the room through nasal or mouth  exhalation.  The exhaled breath represents a jet flow source that takes a conical form and expands spatially till reaching almost zero velocity at a distance $x_{\mathrm{b,max}}$ from the infected person \cite{tang2013airflow}, as illustrated in Fig. \ref{fig:sys_model}. The breathing conical profile expands over the space and across the time with a reduced velocity. The model adopted herein is, in fact, aligned with the experimental findings of reference \cite{tang2013airflow}. More specifically, Fig.~4 in \cite{tang2013airflow} shows the distance and velocity of the exhaled profile along with the breathing expansion area of 30 volunteers. Fig.~4 in \cite{tang2013airflow} infers that, based on the velocity and propagation distance of the exhaled conical profile, the propagation velocity vanishes after a specific distance $x_{\mathrm{b},\max}$ and time. The aerosol of exhaled particles is, therefore, transferred only by the diffusion mechanism.  Thus, the exhaled particles contain pathogens that propagate in the entire room through diffusion process from a continuous circular surface source, which justifies the rationale of adopting the conical~pattern shown in Fig.~\ref{fig:sys_model} above.

	
	The paper assumes that the room boundaries have specific capabilities of full or partial reflection/absorption. While such features of the room can be designed to control infection or support detection using a separate optimizing module, such optimization falls outside the scope of the current paper and is left as a future research direction. At the receiver side, on the other hand, the paper assumes the existence of a single bio-detector whose electrical characteristics change in a way that matches the detection of the virus under study. In particular, the detector is connected to an air sampler that collects the air based on its  volume, which is called the sampling volume and for a specific duration that is called the  sampling time  $t_{\mathrm{s}}$.
	
	One of the paper goals is to analyze the emission profile of a continuous circular source that diffuses along the room. We, therefore, start by characterizing the spatial-temporal viral concentration due to a single viral point source. Afterwards, we use the derived viral concentration of the instantaneous point source to analyze the emission profile of the continuous circular source.
	
	\section{Spatial Temporal Viral Concentration of a Point Source}
	This section analyzes the instantaneous virus concentration in a bounded room due to a single viral point source, similar to the model illustrated in Fig.~\ref{fig:sys_model}. The section first describes the generic solution of the problem by casting its associated partial differential equation (PDE) as a Sturm-Liouville problem. The section then provides a discussion on the provided solution as a function of its eigenvalues' distributions.
	\subsection{Instantaneous Virus Concentration}
	We assume a point source located at $(x_{\mathrm{p}}, y_{\mathrm{p}}, z_{\mathrm{p}})$  with a viral concentration of $Q_{\mathrm{p}} \;  \mathrm{mg/m^3}$. Let $\mathcal{C}_{\mathrm{p}} \left( x, y, z, t \right) \,  \mathrm{mg/m^3}$ be the spatial-temporal viral concentration at time $t$ and at any generic point $(x,y,z)$ inside the bounded room. The mathematical performance of $\mathcal{C}_{\mathrm{p}} \left( x, y, z, t \right) \,  \mathrm{mg/m^3}$ can then be found based on the mass conservation model and Fick's law using the following partial differential equation (PDE)  \cite{arya1999air}:
	\begin{equation} \label{G_PDE}
	\frac{\partial \mathcal{C}_{\mathrm{p}}}{\partial t}  = K_x  \frac{\partial ^2 \mathcal{C}_{\mathrm{p}}}{\partial x^2} +   K_y\frac{\partial ^2 \mathcal{C}_{\mathrm{p}}}{\partial y^2} +  K_z \frac{\partial ^2 \mathcal{C}_{\mathrm{p}}}{\partial z^2}   + \mathcal{S}  + \mathcal{R},
	\end{equation}
	where the $K_\nu  \; (\mathrm{m^2/s})$'s, i.e., the values of the molecular diffusivity along the $\nu$-dimension with $\nu=x,y,z$, respectively, are constants that depend on the diffusive mass type and the surrounding fluid, $\mathcal{R}$ $\mathrm{mg/s.m^3} $ is the concentration rate change due to any reactions, and $\mathcal{S}$ $\mathrm{mg/s.m^3}$ represents any sources and/or sinks that can change the concentration rate. Throughout this paper,  we do not assume any reactions could occur in the room, i.e., $\mathcal{R}=0$. Since we study the concentration for an instantaneous  point source that is released at $t=t_0$ in this section, then $\mathcal{S} = Q_{\mathrm{p}}  \delta\left(t-t_ 0\right)   \delta\left(x-x_\mathrm{p}\right)     \delta\left(y-y_\mathrm{p}\right) $ $\delta\left( z-z_\mathrm{p} \right)$. Thus, we can express $\mathcal{C}_{\mathrm{p}} \left( x,y,z, t \right)$ for different time intervals as follows
	\begin{enumerate}
		\item For $t<t_0$, $\mathcal{C}_{\mathrm{p}} = 0.$
		\item For $t=t_0, \, \mathcal{C}_{\mathrm{p}} = Q_{\mathrm{p}}  \delta\left(x-x_\mathrm{p}\right)     \delta\left(y-y_\mathrm{p}\right) $ $\delta\left( z-z_\mathrm{p} \right)$, where $Q_{\mathrm{p}}$ is the viral emission rate that occurs only at $t =t_0 $.
		\item For $t>t_0$, $\mathcal{C}_{\mathrm{p}}$ is found using a simplified form of the PDE presented in (\ref{G_PDE}):
		\begin{equation} \label{main_pde1}
		\frac{\partial \mathcal{C}_{\mathrm{p}}}{\partial t} =  K_{x}\frac{\partial^2 \mathcal{C}_{\mathrm{p}}}{\partial x^2} + K_{y}\frac{\partial^2 \mathcal{C}_{\mathrm{p}}}{\partial y^2} + K_{z}\frac{\partial^2 \mathcal{C}_{\mathrm{p}}}{\partial z^2}.
		\end{equation}
	\end{enumerate}
	
	To solve \eqref{main_pde1}, we need to consider the following system initial condition (IC) and boundary conditions (BCs):
	\begin{enumerate}
		\item The initial viral concentration is
		\begin{equation} 	\label{main_IC}
		\mathcal{C}_{\mathrm{p}} \left( x, y, z, t_0 \right) =  Q_{\mathrm{p}} \, \delta\left(x-x_\mathrm{p}\right)  \delta\left(y-y_\mathrm{p}\right) \delta\left( z-z_\mathrm{p} \right).
		\end{equation}
		\item The room is bounded in the $\nu$-th direction, where $0 \leq \nu \leq L_\nu$, for $\nu\in\{x,y,z\}$.
		\item Each room side exhibits different absorption and reflection characteristics, which is modeled mathematically as \cite[Ch. 18, Eqn. (18.92)]{seinfeld2016atmospheric},
		\begin{align} \label{main_BC}
		K_{\nu}\frac{\partial \mathcal{C}_{\mathrm{p}} \left(x, y, z, t\right) }{\partial \nu}= d_{\nu_i}\mathcal{C}_{\mathrm{p}} \left(x, y, z, t\right),   \quad  \; \nu= \nu_i,
		\end{align}
		where $i= 1, 2, \, \nu_1= 0, \; \nu_2= L_\nu$, and $d_{\nu_i}$ is the deposition velocity in the $\nu-\mathrm{th}$ direction. In (\ref{main_BC}), $d_{\nu_i}$ is used to represent the absorption/reflection level of a certain surface or wall \cite[Ch. 18]{seinfeld2016atmospheric}, where $d_{\nu_i} = 0$ denotes total reflection scenario. As $d_{\nu_i} $ increases, however, the surface absorption capability increases; a total absorption scenario occurs when $d_{\nu_i} \to \infty$. The BC \eqref{main_BC} is known  mathematically as Robin BC in differential equations  \cite{gockenbach2010pde}. In the following, we rewrite \eqref{main_BC} in a simpler way as
		\begin{equation} \label{BC_subproblem}
		\frac{\partial \mathcal{C}_{\mathrm{p}} }{\partial \nu}= \beta_{\nu_i} \mathcal{C}_{\mathrm{p}},\quad\text{where}~  \beta_{\nu_i}=\frac{d_{\nu_i}}{K_{\nu}}.
		\end{equation}
	\end{enumerate}

	Now, we express $\mathcal{C}_{\mathrm{p}} \left(x, y, z, t\right)$ as a multiplication of independent spatial functions. Such operation is mathematically possible since the diffusion in a specific dimension does not depend on other spatial dimensions \cite{seinfeld2016atmospheric}. More specifically, $\mathcal{C}_{\mathrm{p}} \left(x, y, z, t\right)$ can be written as:
	\begin{equation}  \label{C_sep}
	\mathcal{C}_{\mathrm{p}}\left(x, y, z, t\right)= \mathcal{C}_x \left(x, t \right) \mathcal{C}_y \left(y, t \right) \mathcal{C}_z\left(z, t \right).
	\end{equation}
	Then, we use \eqref{C_sep} to find the equivalent PDE in the $\nu-\mathrm{th}$ dimension from the main PDE \eqref{main_pde1}. We get the following PDE\footnote{The proof of (\ref{sub_PDE}) is provided in Appendix \ref{Appendix_A}.}:
	\begin{equation} \label{sub_PDE}
	\frac{\partial \mathcal{C}_\nu   \left( \nu, t \right)}{\partial t} = K_\nu\frac{\partial^2 \mathcal{C}_\nu \left( \nu, t \right) }{\partial \nu^2}.
	\end{equation}
	As for the IC, we use \eqref{C_sep} in main IC equation \eqref{main_IC} to get:
	\begin{equation} \label{IC}
	\mathcal{C}_\nu(\nu, t_0) = Q_\nu \delta(\nu- \nu_{\mathrm{p}}),
	\end{equation}
	where $Q_\nu$ is the virus concentration across the $\nu-\mathrm{th}$ dimension with $Q_{\mathrm{p}} = Q_x Q_y Q_z$. As for the BCs, a simple manipulation of equations \eqref{C_sep} and \eqref{BC_subproblem} gives the following simplified BCs:
	\begin{align} \label{BC1}
	\frac{\partial \mathcal{C}_\nu}{\partial \nu}= \beta_{\nu_1}\mathcal{C}_\nu, & \; \; \; \nu= 0, \\
	\label{BC2}
	\frac{\partial \mathcal{C}_\nu}{\partial \nu}= \beta_{\nu_2}\mathcal{C}_\nu, & \; \; \; \nu= L_\nu.
	\end{align}

	The bounded-value PDE problem \eqref{sub_PDE} with the BCs \eqref{BC1},  \eqref{BC2},  and the IC \eqref{IC} can be classified among the Sturm-Liouville problems, which can be solved using a proper separation of variables  \cite{gockenbach2010pde, zill2016differential, boyce2004elementary, strauss2007partial}. By applying the variables separation  approach to our problem, we  can replace the PDE system with a set of ordinary differential equations (ODE) that are solved subject to the given BCs and IC. More specifically, we first express  $\mathcal{C}_\nu(\nu,t)$ as a multiplication of separable functions as
	\begin{equation}  \label{Cv_PDE}
	\mathcal{C}_\nu\left(\nu, t\right)= \mathcal{V}\left(\nu\right)\mathcal{T}\left(t\right).
	\end{equation}
	Then, we substitute \eqref{Cv_PDE} in \eqref{sub_PDE} and divide both sides by $K_\nu \mathcal{V}\left(\nu\right)\mathcal{T}\left(t\right)$. We obtain the following equalities:
	\begin{equation} \label{ode_qn}
	\frac{1}{K_{\nu}\mathcal{T}\left(t \right)}   \frac{d \mathcal{T} \left(t\right)}{d t} = \frac{1}{\mathcal{V} \left(\nu\right)}\frac{d^2 \mathcal{V} \left(\nu \right )}{d \nu^2} := - \alpha_\nu,
	\end{equation}
	where $\alpha_\nu$ is a constant, since the first equality in \eqref{ode_qn} equates functions of the two independent variables $t$ and $\nu$. Interestingly, the equalities in \eqref{ode_qn} illustrate how the separation of variable approach converts the PDE equation in \eqref{sub_PDE} into two ordinary differential equations (ODEs), that can be written as follows:
	\begin{align} \label{V_ode}
	& \frac{d^2 \mathcal{V} \left(\nu \right )}{d \nu^2} +\alpha_\nu  {\mathcal{V} \left(\nu\right)} = 0,  \\
	&  \frac{d \mathcal{T}(t)}{\mathcal{T}(t)} + \alpha_\nu K_\nu dt = 0. \label{T_ode}
	\end{align}

	In Sturm-Liouville problems, $\alpha_\nu$ is called an eigenvalue, which can take different values that satisfy the BCs \eqref{BC1} and \eqref{BC2}. Each eigenvalue gives a possible eigenfunction solution, which contributes to the final concentration function in a form of a weighted sum of all possible eigenfunctions that are orthogonal, thanks to the Sturm-Liouville problem structure \cite{gockenbach2010pde, zill2016differential, boyce2004elementary, strauss2007partial}. Finally, the weights of different solutions can be found by using the IC \eqref{IC} and  eigenfunctions' orthogonality property.
	
	First, solve the ODE \eqref{T_ode} gives the following solution:
	\begin{equation} \label{T_comp}
	\mathcal{T} \left( t \right) = c_0  e^{- \left| \alpha_\nu \right| K_\nu  t},
	\end{equation}
	where $c_0$ is a constant. Solving the more complicated ODE \eqref{V_ode}, on the other hand, necessitates the investigation of possible eigenvalues that satisfy \eqref{V_ode}, subject to BCs \eqref{BC1} and~\eqref{BC2}.
	
	To best characterize the above solutions, the following subsection studies the eigenvalues distribution, evaluate them and find the corresponding eigenvectors and weights, which would eventually allow us to express $\mathcal{C}_\nu \left( \nu, t \right)$ as
	\begin{equation} \label{General_Cnu}
	\mathcal{C}_\nu \left( \nu, t \right) = \sum_{n} \ell_{\nu, n}\Phi_n\left( \nu, t \right),
	\end{equation}
	where $\Phi_n\left( \nu, t \right)$  is the $n-\mathrm{th}$ eigenfunction that corresponds to the $n-\mathrm{th}$ eigenvalue, $\lambda_{\nu, n}$, and $\ell_{\nu, n}$ is the corresponding weight. It is worthy to emphasize that $\Phi_n\left( \nu, t \right)$ would, by design, jointly have a separable form as in \eqref{Cv_PDE} and simultaneously satisfy the PDE in \eqref{sub_PDE}.
	
	\subsection{Eigenvalues Distributions}
	To study the distributions of the eigenvalues $\alpha_\nu$, we next discuss three possible cases based on the sign of $\alpha_\nu$. More specifically, the first case below assumes a zero eigenvalue, i.e., $\alpha_{\nu} = 0$. Then, we discuss the positive eigenvalues case, i.e., we let $\alpha_{\nu}= \lambda_{\nu}^2$, where $\lambda_{\nu}^2$ is a positive real number. Lastly, we consider the negative eigenvalues case, i.e., we let $\alpha_{\nu} = - \tilde{\lambda}_{\nu}^2$, where  $\tilde{\lambda}_{\nu}^2$ is a positive real number.
	\subsubsection{Zero Eigenvalue Case}
	First, we consider the ODE in \eqref{V_ode}, with $\alpha_\nu = 0$, which gives the following solution
	\begin{equation} \label{zero_EV}
	\begin{split}
	\mathcal{V}\left( \nu \right) & = c_1 \nu + c_2,
	\end{split}
	\end{equation}
	where $c_1$ and $c_2$ are constants computed to satisfy the BCs. By considering the BCs  \eqref{BC1} and  \eqref{BC2}, we obtain
	\begin{equation} \label{zero_EV_eqns}
	\begin{split}
	c_1 &= \beta_{\nu_1} c_2, \\
	\left(1 - \beta_{\nu_2} L_\nu \right) c_1 &= \beta_{\nu_2} \left( c_1 L_\nu + c_2 \right),
	\end{split}
	\end{equation}
	which can have two possible solutions. The first solution is the trivial solution, i.e., $c_1=c_2 = 0$. Such solution, however, implies that $\mathcal{V}\left( \nu \right)$ is zero, which highlights the impact of partial absorption of the wall, a feature of which is the zero-concentration at steady state. The second solution occurs by substituting the value of $c_1$ as $c_1 = \beta_{\nu_1} c_2$ in $\left(1 - \beta_{\nu_2} L_\nu \right) c_1= \beta_{\nu_2} \left( c_1 L_\nu + c_2 \right)$. A direct inspection of such substitution concludes that, in this case, \eqref{zero_EV_eqns} would be satisfied when $\beta_{\nu_2} = {\beta_{\nu_1}}/ \left({1+ \beta_{\nu_1} L_{\nu}}\right)$, provided that $c_1 = \beta_{\nu_1} c_2$. Such solution, however, represents the existence of a non-zero steady state solution. Under the assumption of partial absorption, the viral concentration due to an instantaneous point source must vanish at steady state, and so the second solution above is indeed unfeasible. In summary, when $\alpha_\nu = 0$, the only acceptable solution is $c_1=c_2 = 0$, which rather means the nonexistence of a zero eigenvalue case.
	
	\subsubsection{Positive Eigenvalue Case}
	To study the positive eigenvalues distribution, we consider $\alpha = \lambda^2$ in  \eqref{V_ode}, i.e.:
	\begin{equation} \label{pEV_ODE}
	\frac{d^2  \mathcal{V} \left(\nu \right)}{d \nu^2} =- \lambda^2_\nu \mathcal{V} \left(\nu \right).
	\end{equation}
	
	The solution of \eqref{pEV_ODE} has a trigonometric periodic function form that can be expressed as follows:
	\begin{equation} \label{V_comp}
	\mathcal{V}\left( \nu \right)= c_3 \cos(\lambda_\nu \nu) + c_4 \sin(\lambda_\nu \nu),
	\end{equation}
	where $c_3$ and $c_4$ are constants that are found from the BCs. Firstly, we use the BC in \eqref{BC1} and obtain $c_4 =  {\beta_{\nu_1}c_3}/{\lambda_\nu}$.  Then, we use the second BC in \eqref{BC2} and find the following relation
	\begin{equation} \label{lambda_positive}
	\tan \left( \lambda_\nu L_\nu  \right) = \frac{\lambda_\nu \left(   \beta_{\nu_1}  - \beta_{\nu_2}\right)} {\beta_{\nu_1} \beta_{\nu_2} + \lambda^2_\nu},
	\end{equation}
	which is used to find different values for $\lambda_\nu$ that satisfy the ODE of $\mathcal{V}\left( \nu \right)$ in \eqref{pEV_ODE}.
	
\begin{figure}[!t]
		\centering
		\includegraphics[width=4.5in]{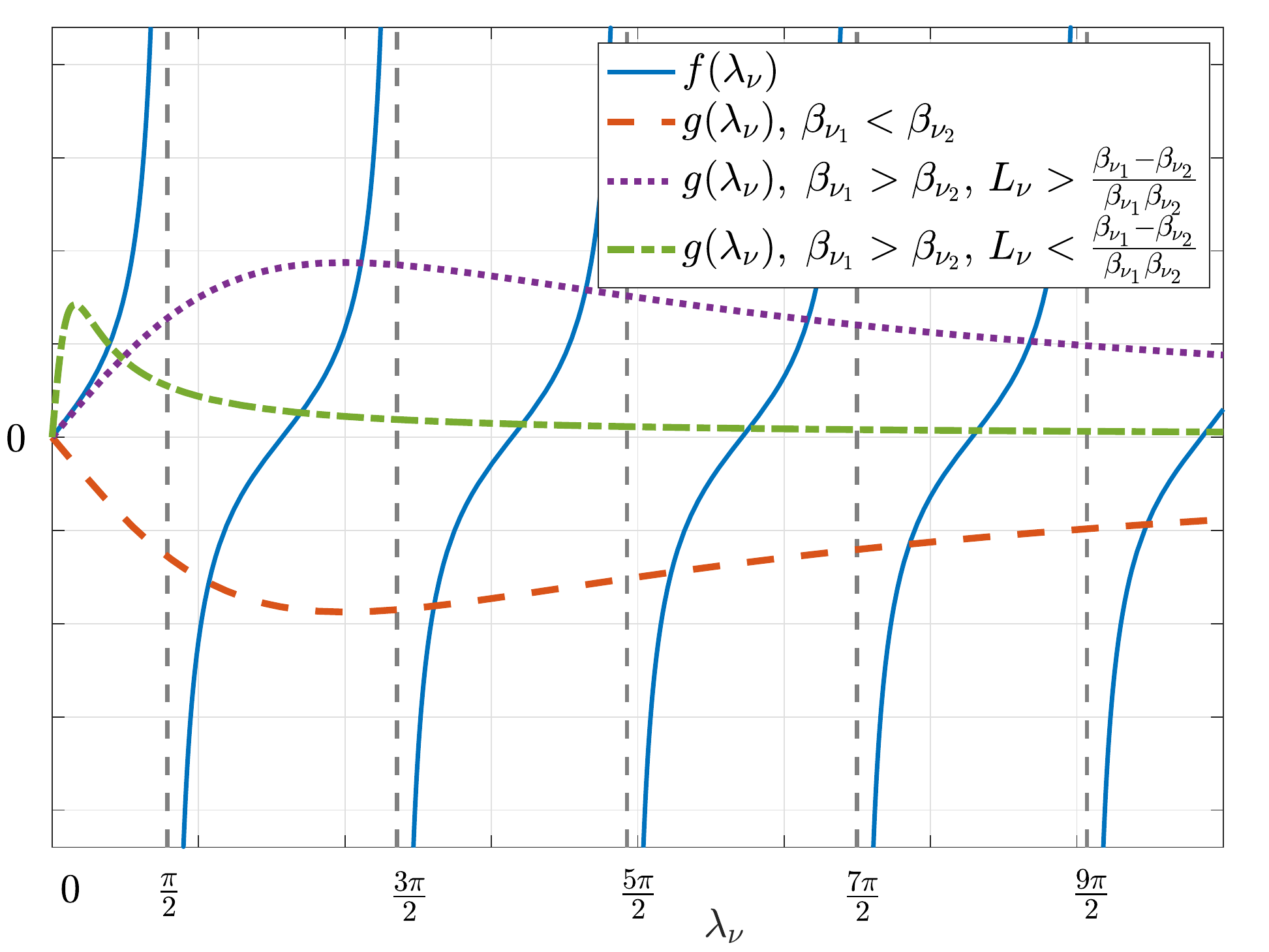}
		\caption{Positive eigenvalues distribution.} \label{fig:pv_EV}
	\end{figure}

	To study the eigenvalues distributions of \eqref{lambda_positive}, we examine both sides of the equality in \eqref{lambda_positive}. To this end, define $f\left( \lambda_\nu \right) = \tan \left( \lambda_\nu L_\nu  \right) $ and $g\left( \lambda_\nu \right) = \frac{\lambda_\nu \left(   \beta_{\nu_1}  - \beta_{\nu_2}\right)} {\beta_{\nu_1} \beta_{\nu_2} + \lambda^2_\nu}$. Fig.~\ref{fig:pv_EV} plots $f\left( \lambda_\nu \right)$ and $g\left( \lambda_\nu \right)$ versus $\lambda_\nu$ so as to graphically characterize the roots of equation \eqref{lambda_positive}. Firstly, we consider the case where $\beta_{\nu_1} < \beta_{\nu_2}$. The eigenvalues in this case are, therefore, at the intersection of $f(\lambda_\nu)$, i.e. the tan function, and $g(\lambda_\nu)$. Such intersection herein is in fact an infinite set of points, each belonging to one interval $( {(k -\frac{1}{2})\pi} , \; k\pi ) $, for one particular integer $k$, as shown in Fig. \ref{fig:pv_EV}.
	
	Secondly, we consider the case where $\beta_{\nu_1} > \beta_{\nu_2}$. In this case, the intersection set is also infinite; however, the eigenvalues belong to intervals $(k\pi, \; {(k +\frac{1}{2})\pi} )$, as shown in Fig. \ref{fig:pv_EV}. At the special case $k=0$, there is an eigenvalue in the interval $(0, \; { \frac{\pi}{2} })$  , if $\frac{df(0)}{d\lambda_\nu} > \frac{dg(0)}{d\lambda_\nu}$, which is equivalent to $L_\nu < \frac{\beta_{\nu_1} - \beta_{\nu_2}}{\beta_{\nu_1}\beta_{\nu_2}}$, as illustrated in Fig. \ref{fig:pv_EV}.
	
	After computing the eigenvalues, i.e., $\lambda_{\nu,n}$, we find the corresponding eigenvectors, i.e., $ \Phi_n\left( \nu, t \right)$,  based on \eqref{Cv_PDE},  \eqref{T_comp} and \eqref{V_comp} as follows:
	\begin{equation}  \label{pv_EFun}
	\Phi_n\left( \nu, t \right)\! = \! \left( \! \cos(\lambda_{\nu,n} \nu) +\frac{\beta_{\nu_1}}{\lambda_{\nu,n}} \sin(\lambda_{\nu,n} \nu)\! \right)e^{-K_\nu \lambda^2_{\nu,n} t}.
	\end{equation}

	The corresponding weight of the eigenvector $ \Phi_n\left( \nu, t \right)$, i.e., ${\ell}_{\nu,n}$, can now be evaluated by using the orthogonality property of eigenvectors and the IC \eqref{IC} \cite{gockenbach2010pde, zill2016differential, boyce2004elementary, strauss2007partial}:
	\begin{equation} \label{weight}
	{\ell}_{\nu,n} = \frac{\int_0^{L_\nu}  \mathcal{C}_{\nu} \left(  \nu, t_0  \right)   \Phi_n \left(  \nu, t_0  \right) d \nu}{  \int_0^{L_\nu}  \Phi^2_n \left(  \nu, t_0  \right)   d \nu }.
	\end{equation}
	Once the integrations of the numerator and the denominator of \eqref{weight} are found\footnote{We show all the computation details in Appendix B.}, equation ${\ell}_{\nu,n} $ can be written afterwards as follows:
	\begin{equation}\label{weight_positive}
	\!\! {\ell}_{\nu,n} \! \!=\!\! \frac{  4 \lambda_{\nu, n}^3 Q_\nu  \left(   \cos(\lambda_{\nu, n} \nu_\mathrm{p}) + \frac{\beta_{\nu_1}}{\lambda_{\nu, n}} \sin(\lambda_{\nu, n} \nu_\mathrm{p}) \right)\mathrm{e}^{K_\nu \lambda_{\nu, n}^2 t_0} }{ (\lambda_{\nu, n}^2 \!-\! \beta_{\nu_1}^2\! ) \sin (2 \lambda_{\nu, n} L_\nu \!) \!-\! 2\lambda_{\nu, n} \beta_{\nu_1} \cos(2 \lambda_{\nu, n} L_{\nu}\!) \!+\!\! \rho_{\nu,n}}\!,
	\end{equation}
	where $\rho_{\nu,n} = 2\lambda_{\nu, n} \left(  \left(\lambda_{\nu, n}^2+\beta_{\nu_1}^2\right) L_\nu +\beta_{\nu_1}  \right) $.

	\subsubsection{Negative Eigenvalue Case}
	In the negative eigenvalues case, i.e., when $\alpha_\nu = - \tilde{\lambda}^2_\nu$,  equation \eqref{V_ode} is expressed as:
	\begin{equation} \label{nEV_PDE}
	\frac{d^2\mathcal{V} \left(\nu \right)}{d \nu^2}= \tilde{\lambda}^2_\nu \mathcal{V} \left(\nu \right).
	\end{equation}
	The solution of \eqref{nEV_PDE} can be written either as in hyperbolic form, or in an equivalent exponential form. In the following, we choose to adopt the hyperbolic representation so as to have a similar form to the one presented in the positive eigenvalue case above. More specifically, $ \mathcal{V}\left( \nu \right)$ can be written as follows:
	\begin{equation} \label{V_comp_n}
	\mathcal{V}\left( \nu \right)= c_5 \cosh(\tilde{\lambda}_\nu \nu) + c_6 \sinh(\tilde{\lambda}_\nu \nu),
	\end{equation}
	where $c_5$ and $c_6$ are constants that can be found from BCs \eqref{BC1} and  \eqref{BC2}. The first BC,  \eqref{BC1}, results in $c_9 = {\beta_{\nu_1}}/{\tilde{\lambda}_\nu}$, while the second BC, \eqref{BC2}, yields
	\begin{equation} \label{lambda_negative}
	\tanh \left(\tilde{\lambda}_\nu L_\nu  \right) = \frac{\tilde{\lambda}_\nu \left(   \beta_{\nu_1} - \beta_{\nu_2}  \right)} {\beta_{\nu_1} \beta_{\nu_2} - \tilde{\lambda}^2_\nu  },
	\end{equation}
	which is used to find different values for $\tilde{\lambda}_\nu$ that satisfy the PDE of $\mathcal{V}\left( \nu \right)$ in \eqref{nEV_PDE}.
	
\begin{figure}[!t]
		\centering
		\includegraphics[width=4.5in]{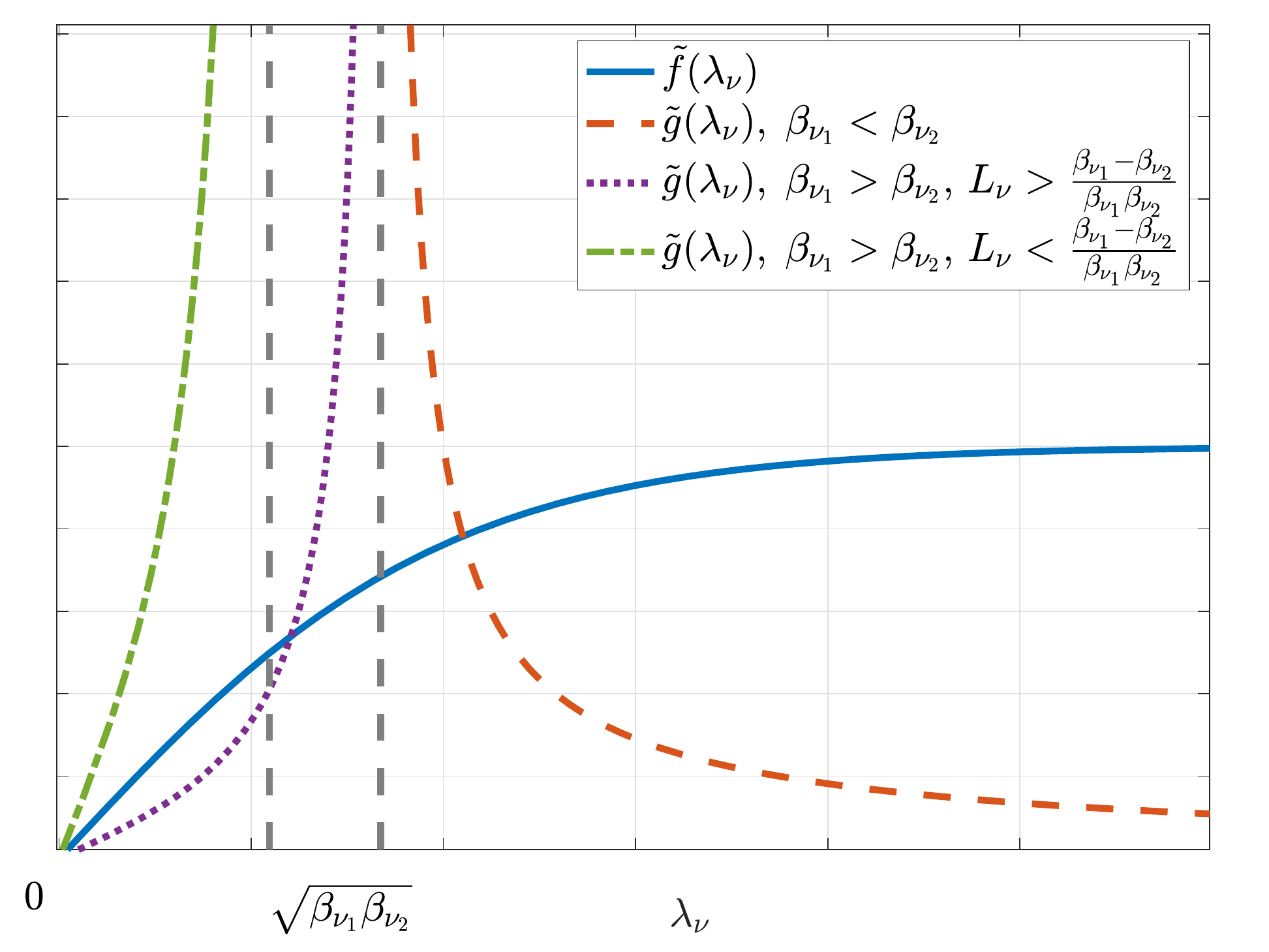}
		\caption{Negative eigenvalues distribution.} \label{fig:nv_EV}
	\end{figure}

	Similar to the discussion of the positive eigenvalues case, we next characterize the negative eigenvalues distributions. Define $\tilde{f}( \tilde{\lambda}_\nu ) = \tanh  ( \tilde{\lambda}_\nu L_\nu )$ and $\tilde{g} ( \tilde{\lambda}_\nu  ) = \frac{\tilde{\lambda}_\nu  (   \beta_{\nu_1} - \beta_{\nu_2}  )} {\beta_{\nu_1} \beta_{\nu_2} - \tilde{\lambda}^2_\nu  }$. Fig.~\ref{fig:nv_EV} plots $\tilde{f}( \tilde{\lambda}_\nu )$ and $\tilde{g} ( \tilde{\lambda}_\nu  )$ versus $\lambda_\nu$. Firstly, when $\beta_{\nu_1} > \beta_{\nu_2}$, both functions are positive and increasing for $\tilde{\lambda}_\nu \in ( 0, \sqrt{\beta_{\nu_1}\beta_{\nu_2}})$. Furthermore, both functions have a zero value at $\tilde{\lambda}_\nu=0$. Fig.~\ref{fig:pv_EV} shows that while $\tilde{f}(\tilde{\lambda}_\nu)$ is concave, $\tilde{g}(\tilde{\lambda}_\nu)$ is convex. Thus, $\tilde{f}(\tilde{\lambda}_\nu)$ and $\tilde{g}(\tilde{\lambda}_\nu)$ may intersect at one point if $\frac{d\tilde{f}(0)}{d\lambda_\nu} > \frac{d\tilde{g}(0)}{d\lambda_\nu}$, which is equivalent to $L_\nu > \frac{\beta_{\nu_1} - \beta_{\nu_2}}{\beta_{\nu_1}\beta_{\nu_2}}$, as illustrated in Fig.~\ref{fig:nv_EV}.

	Secondly, when $\beta_{\nu_1}< \beta_{\nu_2} $, $\tilde{g}(\tilde{\lambda}_\nu)$ is positive if $\tilde{\lambda}_\nu \in (\sqrt{\beta_{\nu_1}\beta_{\nu_2}}, \; \, \infty)$. Since $\tilde{g}(\tilde{\lambda}_\nu)$ decreases and $\tilde{g}(\tilde{\lambda}_\nu)$ increases over $(\sqrt{\beta_{\nu_1}\beta_{\nu_2}}, \; \, \infty)$ as shown in Fig. \ref{fig:nv_EV}, there is always one intersection point (i.e., one feasible eigenvalue).
	
	Based on the above discussion, we conclude that, in the negative eigenvalue case, there is one unique eigenvalue if either $\beta_{\nu_1} < \beta_{\nu_2}$, or  $L_\nu > \frac{\beta_{\nu_1} - \beta_{\nu_2}}{\beta_{\nu_1}\beta_{\nu_2}}$. The corresponding eigenvector, denoted by $\tilde{\Phi} \left( \nu, t \right) $, is then found based on \eqref{Cv_PDE}, \eqref{T_comp} and \eqref{V_comp_n} as follows:
	\begin{equation}  \label{pv_EFun}
	\tilde{\Phi} \left( \nu, t \right) = \left(  \cosh(\tilde{\lambda}_\nu \nu) +\dfrac{\beta_{\nu_1}}{ {\tilde{\lambda}}_{\nu}} \sinh(\tilde{\lambda}_\nu \nu) \right)e^{-K_\nu\tilde{\lambda}^2_{\nu} t}.
	\end{equation}
	The corresponding weight $\tilde{\ell}_{\nu}$ of the eigenvector $\tilde{\Phi} \left( \nu, t \right)$ can then be evaluated based on \eqref{weight} as follows; see Appendix C for more details:
	\begin{equation} \label{weight_n}
	\!\! \tilde{\ell}_{\nu} \! \!=\!\! \frac{  4 \tilde{ \lambda}_{\nu}^3 Q_\nu  \left(   \cosh( \tilde{\lambda}_{\nu} \nu_\mathrm{p}) + \frac{\beta_{\nu_1}}{\tilde{\lambda}_{\nu}} \sinh( \tilde{\lambda}_{\nu} \nu_\mathrm{p}) \right)e^{K_\nu \tilde{\lambda}_{\nu}^2 t_0} }{ (\tilde{\lambda}_{\nu}^2 \!+\! \beta_{\nu_1}^2\! ) \sinh (2 \tilde{\lambda}_{\nu} L_\nu \!) \!+\! 2\tilde{\lambda}_{\nu} \beta_{\nu_1} \cosh(2 \tilde{\lambda}_{\nu} L_{\nu}\!) \!-\!\! \tilde{\rho_{\nu}}}\!,
	\end{equation}
	where $\tilde{\rho}_{\nu} = 2\lambda_{\nu} \left(  \left(\beta_{\nu_1}^2 - \tilde{\lambda}_{\nu}^2\right) L_\nu +\beta_{\nu_1}  \right) $.
	
	\section{Spatial Temporal Viral Concentration of  Exhalation}
	In this section, we utilize the spatial-temporal viral concentration due to an instantaneous point source analyzed in the previous section to characterize the instantaneous virus concentration emitted from a human exhalation in the bounded room,  similar to Fig. \ref{fig:sys_model}. To this end, we consider the exhaled profile as a continuous circular source that diffuses along the room as discussed in section II.
	
	The analysis presented in this section aims at characterizing the spatial-temporal virus concentration due to both continuous (exhaled breath) and temporal sources (sneezing, coughing). We use, therefore, the spatial-temporal viral concentration due to an instantaneous point source analyzed in the previous section so as to study the viral concentration of a continuous emitted circular source, named $\mathcal{C}_{\mathrm{c}} \left( x, y, z, t \right)$, which is assumed to have a uniform virus distribution. We first introduce the following notation to express the spatial-temporal concentration of a point source, $\mathcal{C}_\mathrm{p} \left( x, y, z, t \right)$,
	\begin{equation}
	\mathcal{C}_\mathrm{p} \left( x, y, z, t \right) = \mathcal{C} \left(  x, y, z, t ; x_{\mathrm{p}}, y_{\mathrm{p}}, z_{\mathrm{p}}, t_{0}  \right),
	\end{equation}
	where the right hand side denotes the spatial-temporal concentration due to a point source located at $\left( x_\mathrm{p}, y_{\mathrm{p}}, z_{\mathrm{p}} \right)$ which is released at $t = t_0$.

	Consider the point $\left( x_\mathrm{p}, y_{\mathrm{p}}, z_{\mathrm{p}} \right)$ that is located at the circular plan center, as shown in Fig. \ref{fig:sys_model}. Let $r_c$ be the radius of the circular area. The viral concentration of the continuous emitted circular source $\mathcal{C}_{\mathrm{c}} \left( x, y, z, t \right)$ can then be found  based on $\mathcal{C}_\mathrm{p} \left( x, y, z, t \right)$ as follows:
	\begin{equation} \label{C_c}
	\mathcal{C}_{\mathrm{c}} \left( x, y, z, t \right) \! = \!\! \int_{t_0}^{t}  \!\! \int_{z_1}^{z_2} \! \!\! \int_{y_1 }^{y_2 }  \!\!  \mathcal{C} \left(  x, y, z, t ; x_{\mathrm{p}}, y_0, z_0, \tau \right)    dy_0  dz_0 d\tau,
	\end{equation}
	where $z_1 = z_\mathrm{p} - r_{\mathrm{c}}$, $z_2 = z_\mathrm{p} + r_{\mathrm{c}}$, $y_1 = y_\mathrm{p} - \sqrt{r_{\mathrm{c}}^2 - (z-z_0)^{2}}$, and $y_2 = y_\mathrm{p} + \sqrt{r_{\mathrm{c}}^2 - (z-z_0)^{2}}$. After evaluating the integral of $y_0$,  equation \eqref{C_c} reduces to
	\begin{align} \label{C_c1}
	&\!\!\!\! \mathcal{C}_{\mathrm{c}} \left( x, y, z, t \right)  \!\! =   \!\!  \int_{t_0}^{t}  \!\!  \int_{z_1 }^{z_2 } \!\!   \left(\!  \sum_{n}  \! \widehat{{\ell}}_{y,n} \left(  \tau  \right)  \Phi_n \left( y, t \right)  + \widehat{\tilde{\ell}}_y \left(  \tau  \right) \tilde{\Phi}\left(y,t\right)\!\!  \right)  \!\!   \left(  \sum_{n} \! \ell_{z,n} \left( z_0, \tau \right)  \Phi_n \left(z,t \right)   + \tilde{\ell}_z \left( z_0, \tau \right) \tilde{\Phi}\left(z,t\right)    \right )
	\nonumber \\
	&\!\!\!\!\!\!  \times\!  \left( \! \sum_{n}\!  \ell_{x,n} \left( x_{\mathrm{p}}, \tau \right) \Phi_n\left( x, t \right)  + \tilde{\ell}_x \left( x_{\mathrm{p}}, \tau \right) \tilde{\Phi}\left( x, t \right) \! \right) \! dz_0 d\tau,
	\end{align}
	where $\widehat{{\ell}}_{y,n} \left(  \tau  \right)$ is expressed in terms of ${\ell}_{y,n}$ as
	\begin{equation}\label{l_hat}
	\widehat{{\ell}}_{y,n} \left(  \tau  \right) =  \int_{y_1 }^{y_2 } {\ell}_{y,n} \left( y_0, \tau\right) dy_0,
	\end{equation}
	where ${\ell}_{y,n}(.,.)$ is the weight associated with the positive eigenvalue case, as illustrated in \eqref{weight_positive}. Note that the arguments $y_0$ and $\tau$ of ${\ell}_{y,n}$ in \eqref{l_hat} represent the variables $\nu_p$ and $t_0$ in \eqref{weight_positive}, respectively.  $\widehat{{\ell}}_{y,n} \left(  \tau  \right)$ can now be readily written as:
	\begin{align}
	& \widehat{{\ell}}_{y,n} \left(  \tau  \right) =  4 \lambda_{y, n}^2 e^{ K_y \lambda_{y, n}^2  \tau   }    \frac{       \sin(\lambda_{y, n} y_\mathrm{2}) \! - \! \sin(\lambda_{y, n} y_\mathrm{1})+ \! \frac{\beta_{y_1}}{\lambda_{y, n}} \! \left(   \cos(\lambda_{y, n} y_1) \! -\! \cos(\lambda_{y, n} y_2) \right)  } { (\lambda_{y, n}^2 \!-\! \beta_{y_1}^2\! ) \sin (2 \lambda_{y, n} L_y \!) \!-\! 2\lambda_{y, n} \beta_{y_1} \cos(2 \lambda_{y, n} L_{y}\!) \!+\!\! \rho_{y,n}}.
	\end{align}
	$\widehat{\tilde{\ell}}_y \left(  \tau  \right)$, on the other hand, can be found based on $\tilde{\ell}_y \left( y_0, \tau\right)$ as follows:
	\begin{equation}\label{l_tilde}
	\widehat{\tilde{\ell}}_y \left(  \tau  \right) =  \int_{y_1 }^{y_2 } \tilde{\ell}_y \left( y_0, \tau\right) dy_0,
	\end{equation}
	where $\tilde{\ell}_y \left(.,.\right)$ is the weight associated with the negative eigenvalue case, as illustrated in \eqref{weight_n}. Note that the arguments $y_0$ and $\tau$ of $\tilde{\ell}_y$ in \eqref{l_tilde} represent the variables $\nu_p$ and $t_0$ in \eqref{weight_n}, respectively. $\widehat{\tilde{\ell}}_y \left(  \tau  \right)$ can now be readily written as:
	\begin{align}
	&\widehat{\tilde{\ell}}_y \left(  \tau  \right) =  4\tilde{\lambda}_y^2 e^{K_y \tilde{\lambda}_y^2 \tau }      \frac{        \sinh( \tilde{\lambda}_{y} y_2)  - \sinh( \tilde{\lambda}_{y} y_1) + \frac{\beta_{y_1}}{\tilde{\lambda}_{y}} \left(  \cosh( \tilde{\lambda}_{y} y_2) - \cosh( \tilde{\lambda}_{y} y_1)\right)   }{ (\tilde{\lambda}_{y}^2 \!+\! \beta_{y_1}^2\! )  \sinh (2 \tilde{\lambda}_{y} L_y \!) \!+\! 2\tilde{\lambda}_{y} \beta_{y_1} \cosh(2 \tilde{\lambda}_{y} L_{y}\!) \!-\!\! \tilde{\rho_{y}}}.
	\end{align}

	Evaluating the double integrals in \eqref{C_c1} is a complicated mathematical process due to the circular plan integration which results in dependent integral bounds, and due to the inner multiplication of infinite sum functions. It is, therefore, desirable to find a tractable way to evaluate the concentration in \eqref{C_c1}, or to approximate the spatial temporal concentration across the room. One alternative to partially simplify the mathematical computations in \eqref{C_c} is by approximating the integration evaluation over the circular plane with a square plane. For instance, such approximation can be made possible through replacing the circle with either a smaller or larger square than the circular area. For example, choosing the square with side length of $L_{\mathrm{s}}=2r_{\mathrm{c}}$ results in an upper bound on \eqref{C_c1}. Similarly, choosing a square with a diagonal $2r_{\mathrm{c}}$ gives $L_{\mathrm{s}}=\sqrt{2}r_{\mathrm{c}}$, and results in a lower bound on \eqref{C_c1}.
	
	A more accurate approximation of \eqref{C_c1} can be found by integrating over a square with the same area as the original circle. Such a square has a side length $L_{\mathrm{s}} = \sqrt{\pi}r_{\mathrm{c}}$. In this case, the concentration in \eqref{C_c} due to integrating over a square planar surface can be found using the following integration bounds substitutions: $z_1 = z_\mathrm{p} - L_{\mathrm{s}}/2$, $z_2 = z_\mathrm{p} + L_{\mathrm{s}}/2 $, $y_1 = y_\mathrm{p} - L_{\mathrm{s}}/2$, and $y_2 = y_\mathrm{p} + L_{\mathrm{s}}/2 $. The concentration in \eqref{C_c} then reduces to the following expression:
	\begin{align} \label{Css}
	&\mathcal{C}_{\mathrm{s}} \left( x, y, z, t \right)  \!\! = \!\!  \int_{t_0}^{t}  \!\!  \left( \sum_{n}   \widehat{{\ell}}_{y,n} \left(  \tau  \right)  \Phi_n \left( y, t \right)  + \widehat{\tilde{\ell}}_y \left(  \tau  \right) \tilde{\Phi}\left(y,t\right)\!  \right)     \left( \sum_{n}   \widehat{{\ell}}_{z,n} \left(  \tau  \right)  \Phi_n \left( z, t \right)  + \widehat{\tilde{\ell}}_z \left(  \tau  \right) \tilde{\Phi}\left(z,t\right)  \right)
	\nonumber \\
	&  \times \! \left( \! \sum_{n} \ell_{x,n} \left( x_{\mathrm{p}}, \tau \right) \Phi_n\left( x, t \right)  + \tilde{\ell}_x \left( x_{\mathrm{p}}, \tau \right) \tilde{\Phi}\left( x, t \right) \! \right) \!   d\tau,
	\end{align}
	\noindent where $\mathcal{C}_{\mathrm{s}} \left( x, y, z, t \right)$ denotes the respective approximate concentration. In fact, the numerical results in section VI suggest that that integration simplification in \eqref{Css} yields a tight approximation as compared to the original concentration expression \eqref{C_c1}. The results also illustrate the numerical behavior of both the upper and the lower bounds described above.

	\section{Viral Detection Performance}
	In this section, we aim to study the ability of detecting viruses from exhaled breath in the bounded environment, through the utilization of electronic-based biosensors, such as the Silicon NanoWire (Si-NW) field effect transistor (FET) \cite{shen2011integrating, khalid2018system, khalid2020modeling}. The paper assumes that the virus presence can be detected with the help of the virus antibodies, which afre attached to the Si-NW, which is placed between the FET source and drain. Prior to the biosensor, an aerosol sampler is used to collect the suspended air during a specific sampling time $T_{\mathrm{s}}$ in a specific sampler volume $V_\mathrm{s}$. Among the possible sampler alternatives, we choose to adopt the electrostatic aerosol sampler due to its commercial availability, sensitivity and ability in sampling nano-sized pathogens particles (regardless of the size of pathogen-laden droplets which could be in the micrometer range).
	
	Given the above receiver model, the received virus concentration is then modeled as in \cite{khalid2018system, khalid2020modeling}:
	\begin{equation} \label{received_C}
	\mathcal{C}_\mathrm{r} = \eta \gamma \mathcal{C}_{\mathrm{samp}} + N,
	\end{equation}
	where $\eta$ is the sampling efficiency, $\gamma$ is the probability of virus binding, $ \mathcal{C}_{\mathrm{samp}}$ is the mean of sampled virus concentration, and $N$ is the additive noise that captures the effect of flicker, thermal, interference  and binding noise. In \eqref{C_c1}, the random variable $N$ is modeled as a zero mean Gaussian noise with variance $\sigma^2$. $\mathcal{C}_{\mathrm{samp}}$, on the other hand, is expressed as the integration of viral concentration over  $V_{\mathrm{s}}$  during $T_\mathrm{s}$ as follows:
	\begin{equation} \label{cmean}
	\mathcal{C}_{\mathrm{samp}} = \int_{T_{\mathrm{s}}} \int_{V_{\mathrm{s}}}  \mathcal{C}_{\mathrm{c}} \left( x, y, z, t \right)  dx dy dz dt,
	\end{equation}
	where $\mathcal{C}_{\mathrm{c}} \left( x, y, z, t \right)$ is defined in \eqref{C_c1}. Consider a cuboid sampler with dimensions $a_x$, $a_y$, and $a_z$  that is centered at $x_\mathrm{d}, \, y_\mathrm{d}, \, z_\mathrm{d},$ in the $x$, $y$ and $z$ dimensions, respectively. Thus, a simplified expression of $\mathcal{C}_{\mathrm{samp}}$ can be found by using the approximate concentration of a square area in \eqref{Css} as follows:
	\begin{equation} \label{cmean_s}
	\begin{split}
	\mathcal{C}_{\mathrm{samp}} \! &\approx \! \int_{t-T_{\mathrm{s}}}^t \!  \int_{t_0}^{t_{\mathrm{e}}}  \!\!  \left( \sum_{n}   \widehat{{\ell}}_{y,n} \left(  \tau  \right)  \Psi_{n,y} \left( y  \right)  + \widehat{\tilde{\ell}}_y \left(  \tau  \right) \tilde{\Psi}_y \left( t\right) \! \right)     \left( \sum_{n}   \widehat{{\ell}}_{z,n} \left(  \tau  \right)  \Psi_{n,z} \left(  t \right)  + \widehat{\tilde{\ell}}_z \left(  \tau  \right) \tilde{\Psi}_z\left(t \right) \right)     \\
	& \times  \left( \! \sum_{n} \ell_{x,n} \left( x_{\mathrm{p}}, \tau \right) \Psi_{n,x} \left( t \right)  + \tilde{\ell}_x \left( x_{\mathrm{p}}, \tau \right) \tilde{\Psi}_x \left(  t \right) \! \right)  d\tau dt,  
	\end{split}
	\end{equation}
	where the respective expressions of $\Psi_{n,\nu} \left(t \right)$ and $\tilde{\Psi}_\nu \left(  t \right)$ can be generally written as:
	\begin{align}
	& \Psi_{n,\nu} \left( t \right)  = \Bigg[  \! \sin\left(\lambda_{\nu,n} \left( \nu_\mathrm{d} +\frac{a_\nu}{2}\right)  \right) - \sin \left(\lambda_{\nu,n} \left( \nu_\mathrm{d} -\frac{a_\nu}{2}\right)  \right)   \nonumber \\ &  - \frac{\beta_{\nu_1}}{\lambda_{\nu,n}} \left(  \cos \left( \lambda_{\nu,n} \left( \nu_\mathrm{d} +\frac{a_\nu}{2}\right) \right)  -  \cos\left( \lambda_{\nu,n}  \left( \nu_\mathrm{d} -\frac{a_\nu}{2}\right) \right) \right)  \! \Bigg]   \frac{e^{-K_\nu \lambda^2_{\nu,n} t}}{\lambda_{\nu, n}},
	\end{align}
	and
	\begin{align}
	& \tilde{\Psi}_\nu  \left(  t \right)  = \Bigg[  \! \sinh \left(\tilde{\lambda}_{\nu} \left( \nu_\mathrm{d} +\frac{a_\nu}{2}\right)  \right) - \sinh \left( \tilde{\lambda}_{\nu} \left( \nu_\mathrm{d} -\frac{a_\nu}{2}\right)  \right)   \nonumber \\ &  +\frac{\beta_{\nu_1}}{\tilde{\lambda}_{\nu}} \left(  \cosh \left( \tilde{\lambda}_{\nu} \left( \nu_\mathrm{d} +\frac{a_\nu}{2}\right) \right)  -  \cosh \left( \tilde{\lambda}_{\nu}  \left( \nu_\mathrm{d} -\frac{a_\nu}{2}\right) \right) \right)  \! \Bigg]   \frac{e^{-K_\nu \tilde{\lambda}^2_{\nu} t}}{\tilde{\lambda}_{\nu}}.
	\end{align}
	
	Based on the above expressions, we next assess the detection ability in the proposed setup by analyzing the probability of miss-detection. We start by comparing $\mathcal{C}_\mathrm{samp}$ with the maximum likelihood threshold $\mathcal{C}_\mathrm{th}$ to decide whether the pathogens exist at the room. Denote by $\mathsf{T}$ the event of pathogens existence, and by $\mathsf{F}$ the event of pathogens absence. We then use the following decision rule
	\begin{equation}  \label{MAP}
	\mathcal{P}(\mathsf{F}| \mathcal{C}_\mathrm{r}) \underset{\mathsf{T}}{\overset{\mathsf{F}}{\gtrless}} \mathcal{P}(\mathsf{T}|\mathcal{C}_\mathrm{r}).
	\end{equation}
	Using Bay's rule, and given that the events of detecting and not detecting pathogens at the receiver side are equally likely, the decision rule in \eqref{MAP} boils down to the following maximum-likelihood inequality:
	\begin{equation}
	\mathcal{P} \left(C_\mathrm{r} |\mathsf{F}\right) \underset{\mathsf{T}}{\overset{\mathsf{F}}{\gtrless}} \mathcal{P}(\mathcal{C}_\mathrm{r}|\mathsf{T})
	\label{prob}
	\end{equation}
	Based on the received virus concentration expression \eqref{received_C}, one can rewrite \eqref{prob} as
	\begin{equation}
	\frac{1}{\sqrt{2 \pi \sigma^2}} \mathrm{e}^{\frac{-(\mathcal{C}_\mathrm{r}-\eta\gamma \mathcal{C}_\mathrm{samp})^2}{2\sigma^2}} \underset{\mathsf{T}}{\overset{\mathsf{F}}{\gtrless}}   \frac{1}{\sqrt{2 \pi \sigma^2}} \mathrm{e}^{\frac{-(\mathcal{C}_\mathrm{r})^2}{2\sigma^2}},
	\end{equation}
	which can be simplified as follows:
	\begin{equation}
	\mathcal{C}_\mathrm{r}\underset{\mathsf{T}}{\overset{\mathsf{F}}{\gtrless}} \frac{\eta \gamma \mathcal{C}_\mathrm{samp}}{2}.
	\end{equation}
	Thus, the maximum likelihood threshold $\mathcal{C}_\mathrm{th}$ is defined as
	\begin{equation}
	\mathcal{C}_\mathrm{th}= \frac{\eta \gamma \mathcal{C}_\mathrm{samp}}{2}.
	\end{equation}
	The probability of miss-detection, $\mathcal{P}_\mathrm{md} $, can then be expressed as follows
	\begin{equation}\label{Prob_missed}
	\mathcal{P}_\mathrm{md} = \left(\mathcal{C}_{\mathrm{r}} \leq \mathcal{C}_\mathrm{th}|\mathsf{T}\right)= \mathcal{Q}\left(  \frac{\eta\gamma  \mathcal{C}_\mathrm{samp}}{\sqrt{2^3 \sigma^2}} \right) := \mathcal{Q} \left(\sqrt{ \frac{\Gamma \mathcal{C}^2_{\mathrm{samp}}}{Q_{\mathrm{p}}} }  \right)
	\end{equation}
	where $\mathcal{Q} \left( . \right)$ represents the right tail distribution function of the standard Gaussian distribution, i.e., $\mathcal{Q} \left( x \right) = \frac{1}{2} \int_{x}^{\infty} \mathrm{e}^{-\frac{u^2}{2}}du$,  and where $\Gamma=  Q_{\mathrm{p}} \left( \eta \gamma \right)^2 /  ( { {2^3 \sigma^2}} )$ represents a scaled ratio of the detected virus point source concentration and the noise variance. Observe that the definition of $\Gamma$ resembles, to some extent, the concept of signal-to-noise ratio (SNR) in conventional wireless communications systems \cite{goldsmith2005wireless}. The $\mathcal{Q} \left( . \right)$ expression of the probability of miss-detection in \eqref{Prob_missed} is also akin to the classical communications systems probability of error as a function of SNR. In light of such analogy, we choose to simulate the impact of changing $\Gamma$ on the probability of miss-detection, as illustrated in the next section of the paper.

\section{Numerical Results}
In this section, we present the numerical results that validate the instantaneous spatial viral concentration expressions in a bounded environment and quantify the proposed system capabilities at virus detection using appropriate biosensors. We start by investigating the number of positive eigenvalues needed to characterize the spatial-temporal concentration of a point source accurately. Then, we study the viral concentration performance under different absorption/reflection boundaries characteristics and validate the solution by checking the PDE and BCs. After that, we study the spatial-viral concentration due to exhaled breath and study the derived bounds and approximation expressions.  Finally, we study the probability of miss-detection considering several features of the sampling and detecting receiver and different reflecting/absorbing boundary abilities.  Throughout the numerical results, we use $K_\nu = 2.42\times 10^{-5}\, \mathrm{m^2/s}$  \cite{khalid2020modeling} and choose the deposition velocities based on the average values reported in \cite{gao2007modeling}, which considers different ventilation methods in indoor environments. We also choose $10^5$ positive eigenvalues as a benchmark for the viral concentration bounds and approximation performance.  

\begin{figure}[H]
	\begin{subfigure}[!t]{\columnwidth}
		\centering
		\includegraphics[width=3.9in]{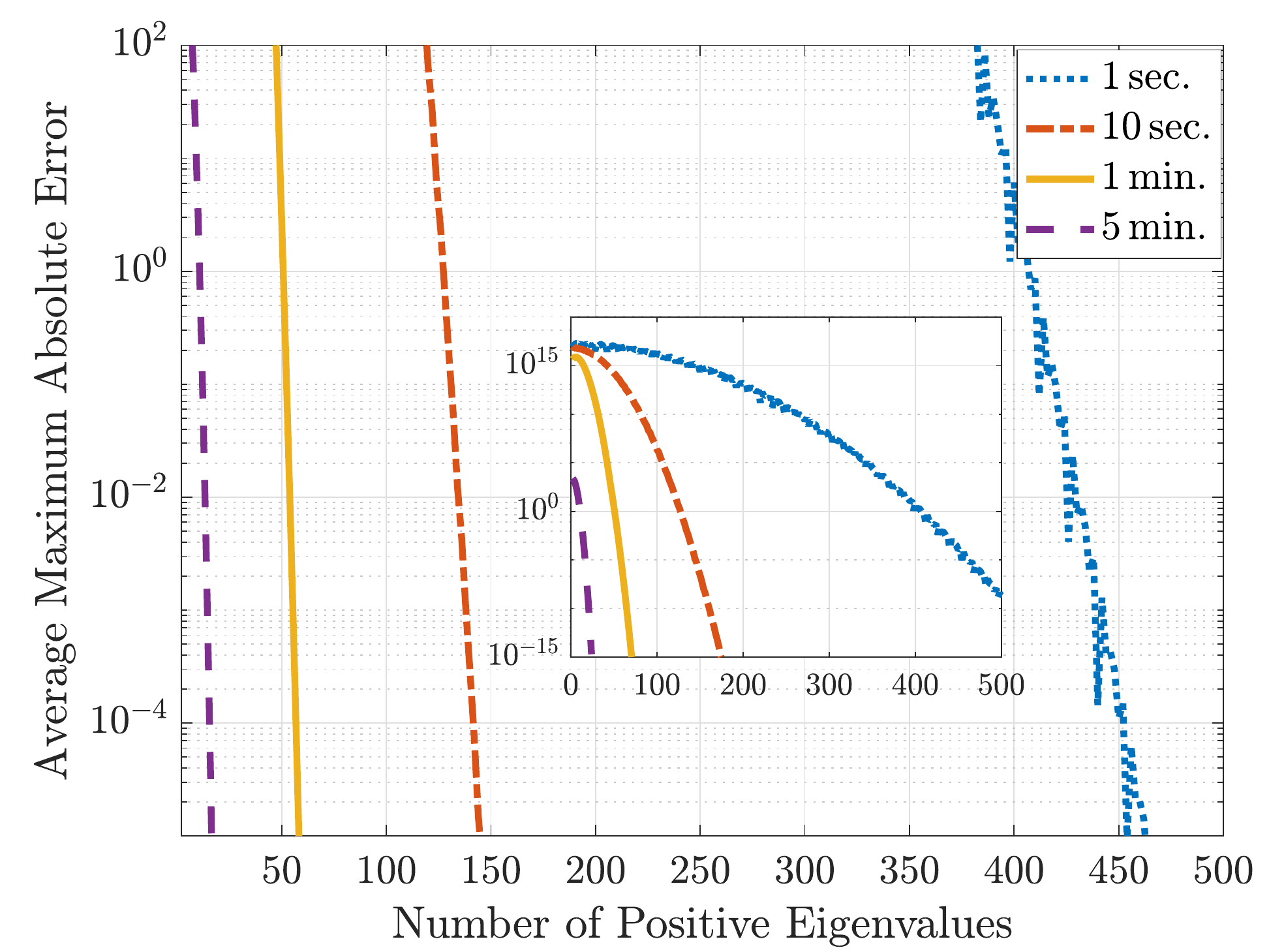}
		\caption{Short-term concentration variations.}
		\label{fig:EV_a}
	\end{subfigure}
	\\
	\begin{subfigure}[!t]{\columnwidth}
		\centering
		\includegraphics[width=3.9in]{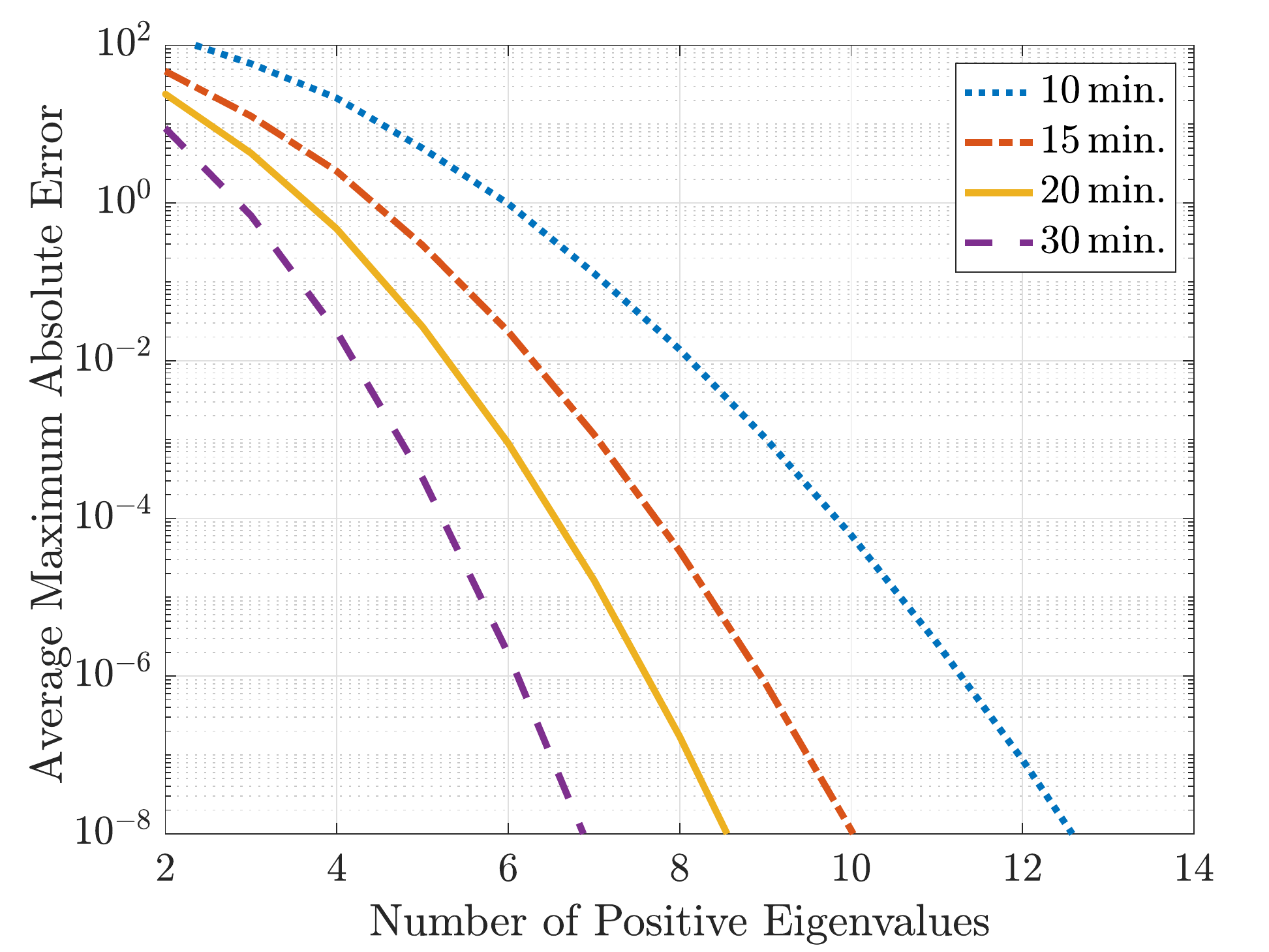}
		\caption{Long-term concentration variations.}
		\label{fig:EV_b}
	\end{subfigure}
	\caption{ The average of maximum absolute errors along the $x-$direction versus different number of positive eigenvalues.}
	\label{fig:EV}
\end{figure}

Firstly, we plot the average maximum absolute error versus a different number of positive eigenvalues in Fig. \ref{fig:EV_a} and Fig. \ref{fig:EV_b}, both under short-time and long-time variations, respectively. To this end, we  study the spatial-temporal concentration due to a point source in $x$-direction with $L_x = 1 \, \mathrm{m}$, and   $x_\mathrm{p} = 0.5\, \mathrm{m}$. Regarding the deposition velocity setting, we assume  one room side is a good reflector with $d_{x_1} = 10^{-7} \mathrm{m/s}$, while the other side has good absorption abilities where $d_{x_2} = 10^{-1}\, \mathrm{m/s}$. For short-time variations, the concentration changes due to an instantaneous point source are expected to be close to pulse shapes. Hence, more terms in the infinite summation are needed to characterize the concentration performance accurately, as also confirmed in Fig. \ref{fig:EV_a}. For example, to achieve a maximum relative error of $10^{-4}$, Fig. \ref{fig:EV_a} shows that approximately 450, 145, and 55 terms are needed to characterize the concentration for $t=1, \, 10, \, \text{and~}60\,  \mathrm{sec.}$, respectively. In the case of the long-term variations, the concentration is expected to change slightly, thus less positive eigenvalue terms are needed to represent the concentration performance, as confirmed in Fig. \ref{fig:EV_b}. Specifically, around 10, 8, 7, and 6 terms are needed to accurately characterize the concentration performance at  $t = 10, \, 15, \, 20, \, \text{and~} 30 \, \mathrm{min.}$, respectively.

\begin{figure}[H]
	\begin{subfigure}[t!]{\columnwidth}
		\centering
		\includegraphics[width=3.5in]{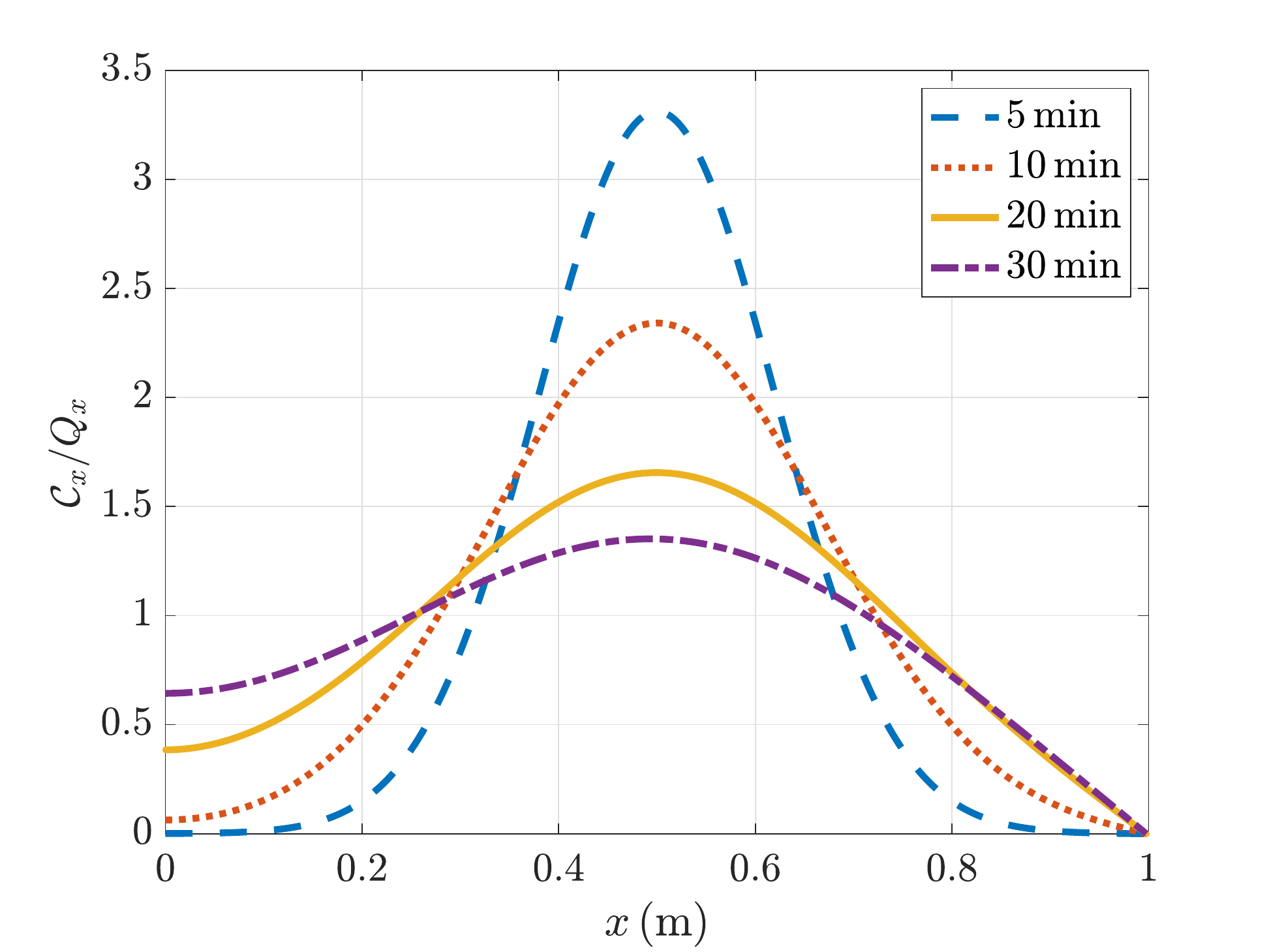}
		\caption{Reflecting and absorbing boundaries with $d_{\nu_1} = 10^{-8} \,  \mathrm{m/s}$ and  $d_{\nu_2} = 10^{-2} \, \mathrm{m/s}$.}
		\label{fig:1dCon_a}
	\end{subfigure}
	\begin{subfigure}[t!]{\columnwidth}
		\centering
		\includegraphics[width=3.5in]{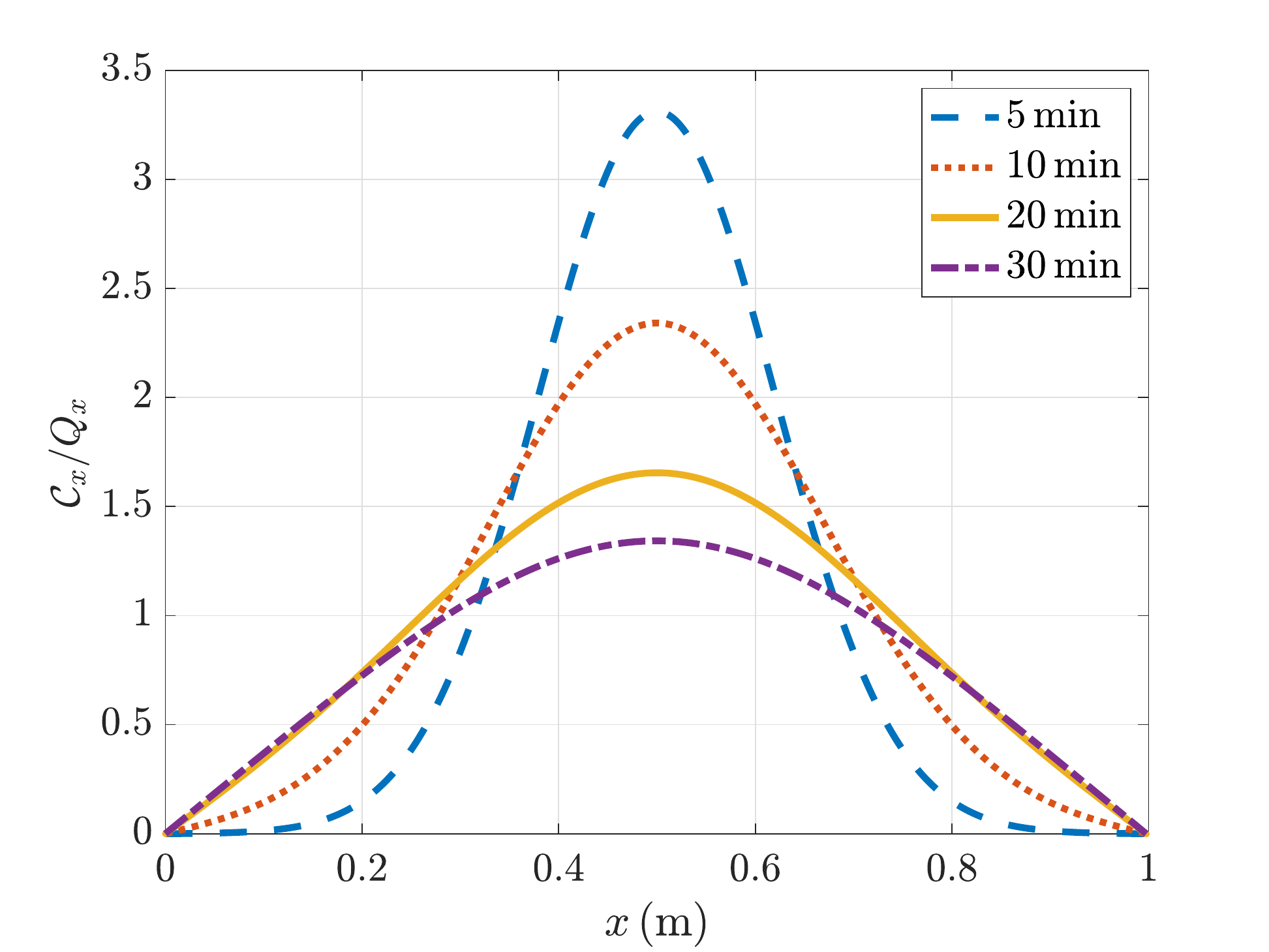}
		\caption{Both absorbing boundaries with $d_{\nu_1} =1  \, \mathrm{m/s}$ and  $d_{\nu_2} =10^{-2} \, \mathrm{m/s}$.}
		\label{fig:1dCon_b}
	\end{subfigure}
	\begin{subfigure}[t!]{\columnwidth}
		\centering
		\includegraphics[width=3.5in]{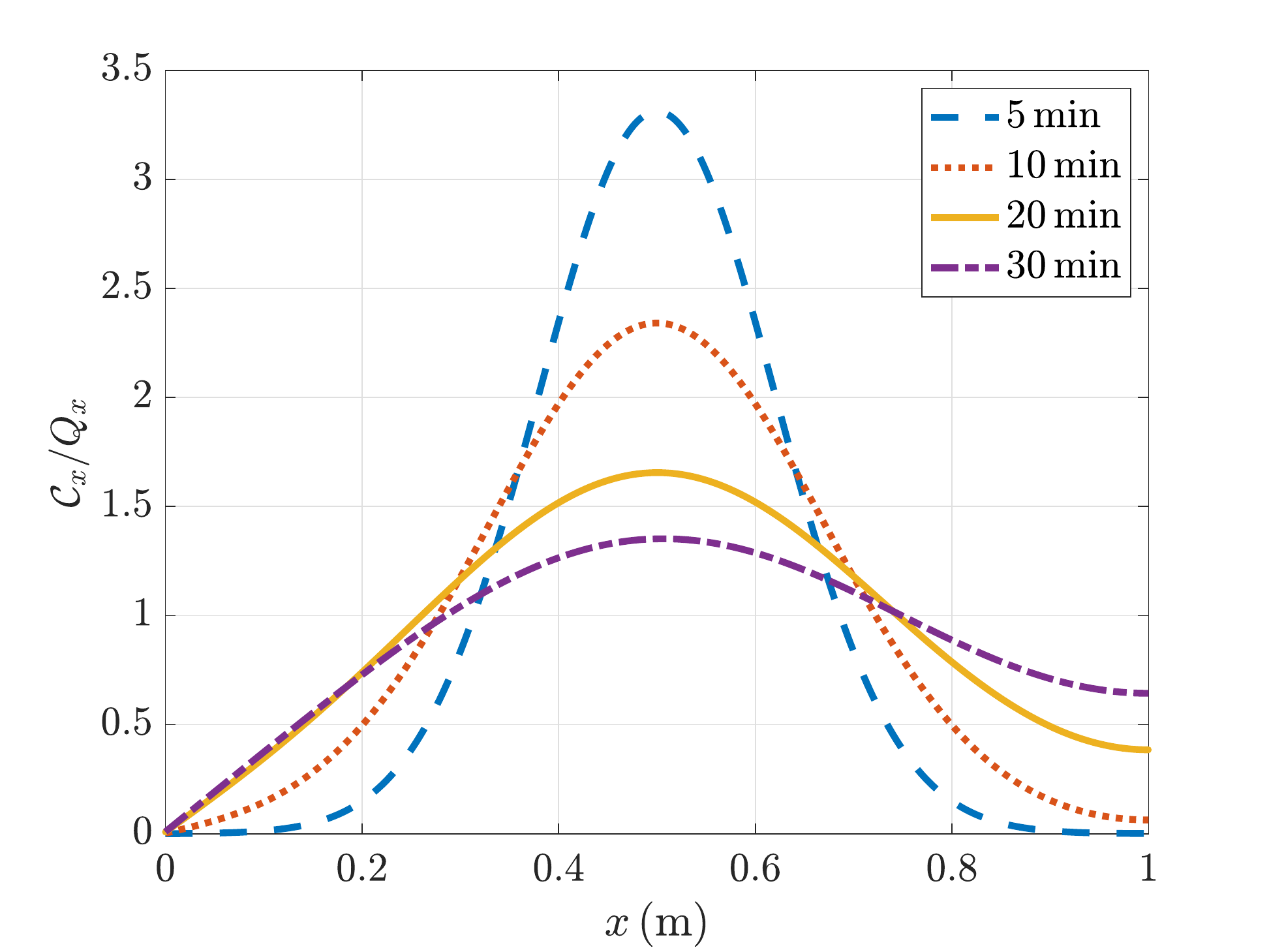}
		\caption{Absorbing and reflecting  boundaries with $d_{\nu_1} = 10^{-2} \, \mathrm{m/s}$ and  $d_{\nu_2} = 10^{-8} \, \mathrm{m/s}$..}
		\label{fig:1dCon_c}
	\end{subfigure}
	\caption{ The impact of different absorption/reflection scenarios on the spatial-temporal viral aerosol concentration due to an instantaneous point source.}
	\label{fig:absorption_reflection}
\end{figure}

Secondly, we study the problem of viral concentration in a bounded environment and validate the problem solution. We focus only on a single-dimension study with an instantaneous point source to easily visualize the effect of partial absorption/reflection and verify the numerical problem solution.  We herein adopt the previous example results shown in Fig. \ref{fig:EV_a} and Fig. \ref{fig:EV_b}, and assume different deposition velocities, as explicitly mentioned in the captions of  Fig. \ref{fig:1dCon_a}, Fig. \ref{fig:1dCon_b} and Fig. \ref{fig:1dCon_c}.  In Fig. \ref{fig:1dCon_a}, we consider the good-reflector good-absorber scenario with $d_{\nu_1} = 10^{-8} \,  \mathrm{m/s}$ and  $d_{\nu_2} = 10^{-2} \, \mathrm{m/s}$, and study the normalized spatial viral concentration with respect to $Q_x$ at different time instances, $t=5,\,10,\,20,\, \text{and~}30\,\mathrm{min.}$. Fig. \ref{fig:1dCon_a} shows that after releasing the instantaneous point source, the concentration decreases slowly around $x_{\mathrm{p}}=0.5\, \mathrm{m}$ as a result of the slow diffusion process that is controlled by $K_x$. On the other hand, the viral concentration increases away from $x_{\mathrm{p}}$, as can be seen in Fig. \ref{fig:1dCon_a}. At the boundary sides, we observe a different behavior due to the deposit velocities variation. On the left side of the room, there is a concentration increase due to the low deposition velocity, which causes viral particle accumulation. On the right side of the room, there is a concentration decrease due to the relatively large deposition velocity, which causes viral particle elimination. In the second scenario, i.e., in Fig. \ref{fig:1dCon_b}, both boundary sides have relatively large deposition velocities. Thus, while the viral particles would vanish at both sides of the room, most of the viral particles remain around $x_{\mathrm{p}}$, as validated in Fig. \ref{fig:1dCon_b}. Finally, in Fig. \ref{fig:1dCon_c}, we show an opposite scenario to the one illustrated in Fig. \ref{fig:1dCon_a}. As expected, Fig. \ref{fig:1dCon_c} shows that the viral particle accumulation occurs near the right side of the room after a relatively larger period of time.

To best verify the validity of the proposed solution, we consider the same scenarios studied in Fig. \ref{fig:absorption_reflection}, and highlight the numerical evaluations of the solutions in the PDE \eqref{sub_PDE}, and the associated BCs in \eqref{BC1} and \eqref{BC2}. The numerical validations of scenarios a, b, and c in Fig. \ref{fig:absorption_reflection} are then listed in Tables \ref{tab:subtable_A}, \ref{tab:subtable_B}, and \ref{tab:subtable_C}, respectively. All three tables illustrate how the evaluation errors are negligible (practically zero) for all scenarios at different time instances, which highlights the validity of the analytical results presented in our paper.

 \begin{table}
 	\caption{Solution verification of Fig. 5 scenarios }
 	\begin{subtable}{\textwidth}
 		\centering
 			\caption{Scenario a}
 		\label{tab:subtable_A}
 		\begin{tabular}[t]{| p{1in} ||p{0.4in}  | p{0.4in} | p{0.4in}  | p{0.4in}  || }
 		  	\hline  \hline
 			Time (min.) & 5   & 10   & 20  & 30    \\ [0.5ex]
 			\hline \hline
 			Maximum absolute error of  PDE \eqref{sub_PDE}	& $1.7\times 10^{-18}$  &$8.7\times 10^{-19}$	 &	$3.8\times 10^{-19}$  & $2.17\times 10^{-19}$  \\
 			\hline
 			Absolute error of  BC  \eqref{BC1} &$9.5\times 10^{-20}$  & $1.7\times 10^{-20}$	 &	$6.8\times 10^{-20}$ &	0 \\
 			 \hline
 			Absolute error of  BC  \eqref{BC2}	& $1.2\times 10^{-12}$ & $7.2\times 10^{-13}$	 &	$3.5\times 10^{-13}$  &	$2.1\times 10^{-13}$  \\
 			\hline \hline
 		\end{tabular}
 	\vspace*{10pt}
 \\
 			\end{subtable}
 	\\
 	\begin{subtable}{\textwidth}
 		\centering
 		 	\caption{Scenario b}
 		\label{tab:subtable_B}
 		\begin{tabular}[t]{| p{1in} ||p{0.4in}  | p{0.4in} | p{0.4in}  | p{0.4in}  || }
 			\hline
 			\hline
 	     	Time (min.) & 5   & 10   & 20  & 30    \\ [0.5ex]
 			\hline
 			Maximum absolute error of  PDE \eqref{sub_PDE}	&  $2.6\times 10^{-18}$  &	$8.7\times 10^{-19}$  	& $3.3\times 10^{-19}$   &	$1.6\times 10^{-19}$  \\
 			\hline
 			Absolute error of  BC  \eqref{BC1}&	$1.6\times 10^{-15}$   &$5.5\times 10^{-16}$ 	 	& $4.4\times 10^{-16}$    & $4.4\times 10^{-16}$   \\
 			\hline
 			Absolute error of  BC  \eqref{BC2}	& $9.2\times 10^{-14}$  &	$2.5\times 10^{-14}$    &	$1.6\times 10^{-13}$  &	$1.8\times 10^{-13}$   \\		
 			\hline \hline
 		\end{tabular}	
 	\vspace*{10pt}
 \\
 		\end{subtable}
 	\\
\begin{subtable}{\textwidth}
	\centering
		\caption{Scenario c}
	\label{tab:subtable_C}
	\begin{tabular}[t]{| p{1in} ||p{0.4in}  | p{0.4in} | p{0.4in}  | p{0.4in}  || }
		\hline
		\hline
		Time (min.) & 5   & 10   & 20  & 30    \\ [0.5ex]
		\hline
		Maximum absolute error of  PDE \eqref{sub_PDE} & $2.6\times 10^{-18}$ 	 	&$8.7\times 10^{-19}$   & $3.3\times 10^{-19}$ 	  &	$1.6\times 10^{-19}$    \\
		\hline
		Absolute error of  BC  \eqref{BC1}&	$4.6\times 10^{-16}$   & $1.1\times 10^{-15}$ 	 &	$4.4\times 10^{-16}$   &0 \\
		\hline
	Absolute error of  BC  \eqref{BC2} &	$7.2\times 10^{-15}$   & $1.4\times 10^{-15}$ 	 &	$5.1\times 10^{-16}$   &  $1.2\times 10^{-16}$ 	  \\
		\hline \hline
	\end{tabular}

\end{subtable}
\end{table}

\begin{figure}[H]
	\begin{subfigure}[b]{\columnwidth}
		\centering
		\includegraphics[width=3.3in]{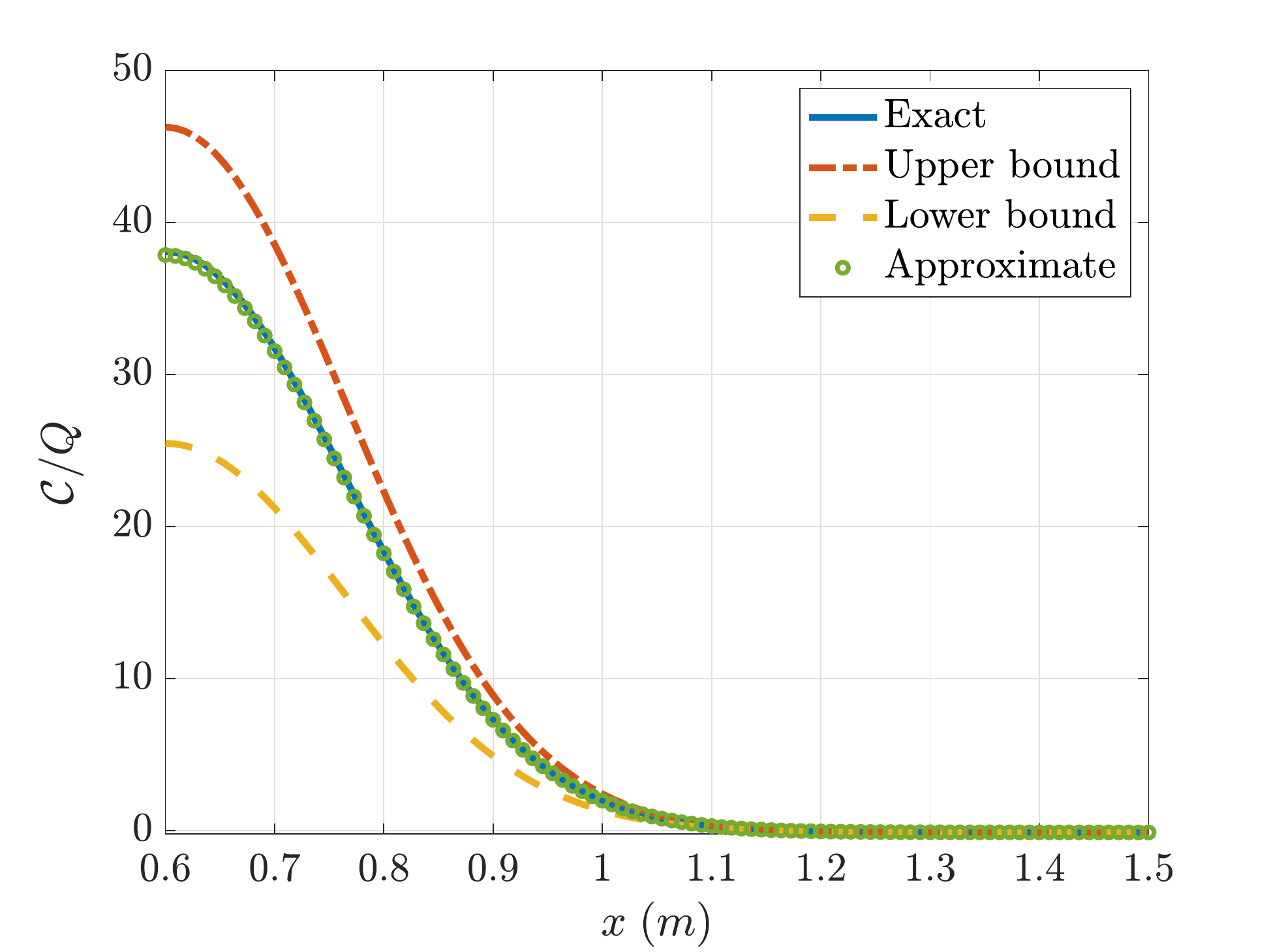}
		\caption{Viral concentration in the $x$-direction at $y=0.4\, \mathrm{m}$ and $z=1.5\, \mathrm{m}$.}
		\label{fig:bounds_x}
	\end{subfigure}
	
	\begin{subfigure}[b]{\columnwidth}
		\centering
		\includegraphics[width=3.3in]{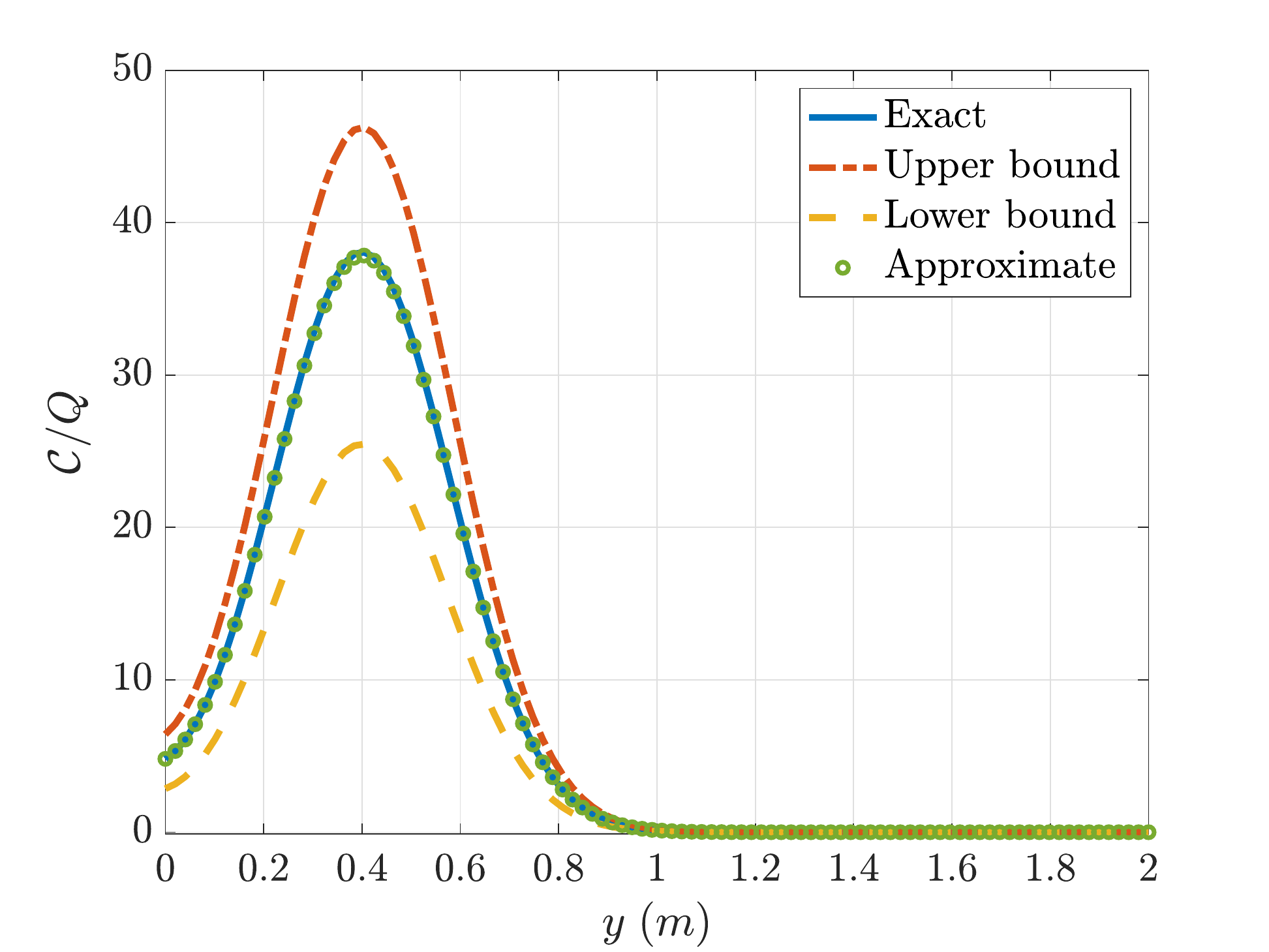}
		\caption{Viral concentration in the $y$-direction at $x=0.6\, \mathrm{m}$ and $z=1.5\, \mathrm{m}$.}
		\label{fig:bounds_y}
	\end{subfigure}
	
	\begin{subfigure}[b]{\columnwidth}
		\centering
		\includegraphics[width=3.3in]{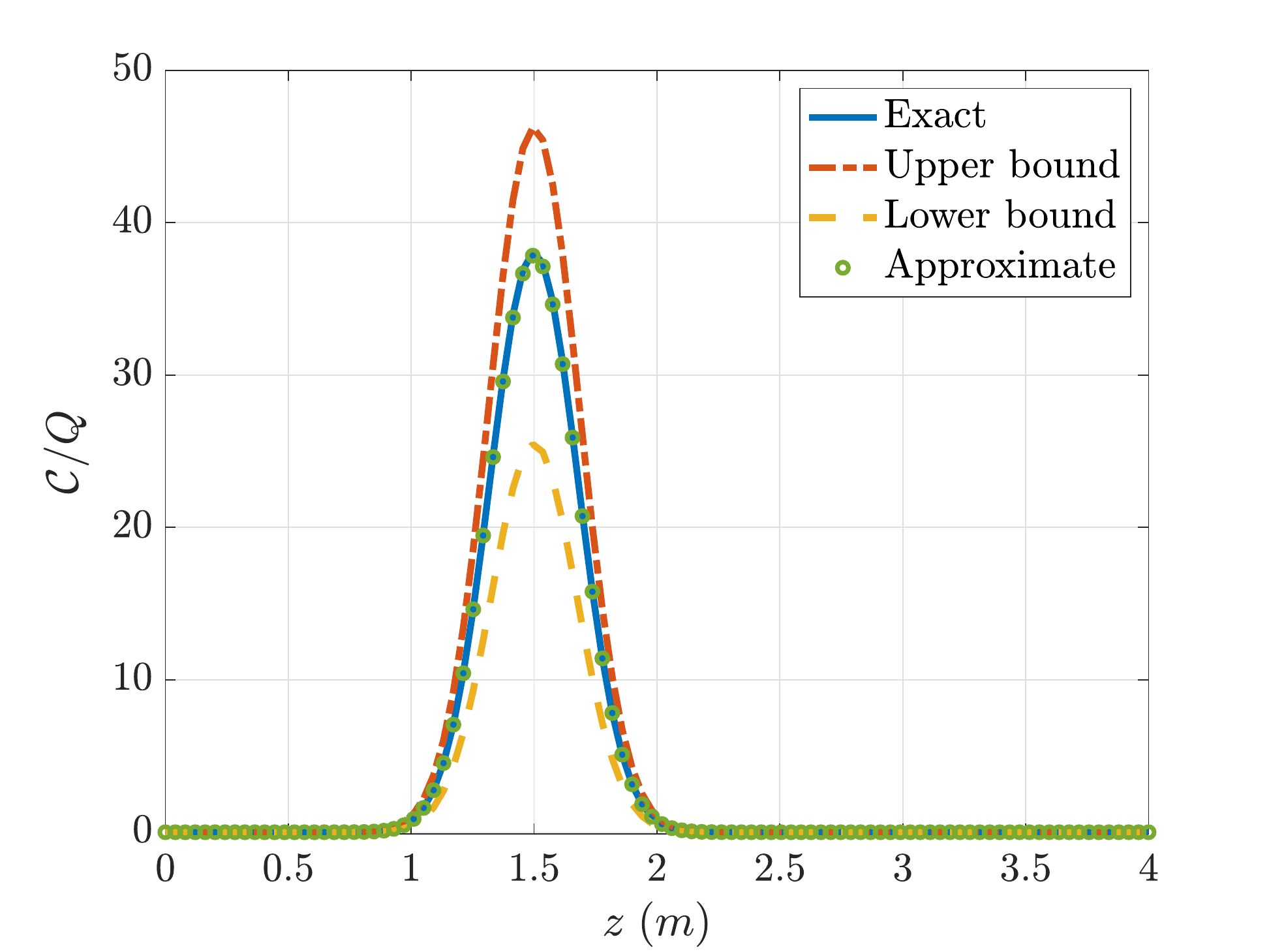}
		\caption{Viral concentration in the $z$-direction at $x=0.6\, \mathrm{m}$ and $y=0.4\, \mathrm{m}$.}
		\label{fig:bounds_z}
	\end{subfigure}
	\caption{ Spatial concentration due to a human exhalation for 1 minute and evaluated after 10 minutes. The circular exhaled planar is located at $x_{\mathrm{p}} =  0.6\, \rm{m}$, $y_{\mathrm{p}} =0.4\, \rm{m}$ and $z_{\mathrm{p}} =  1.5\, \rm{m}$. The spatial concentration is shown versus each dimension and the comparison is provided between the exact, upper bound, lower bound and approximate expressions.}
	\label{fig:Ex3}
\end{figure}

Thirdly, we study the spatial viral concentration in a small bounded environment (e.g., an elevator), due to an infected person that has been in the room for 1 minute. In this scenario, the spatial dimensions are set as follows: $L_x= 1.5\, \mathrm{m}$, $L_y= 3\, \mathrm{m}$, and $L_z= 4\, \mathrm{m}$. For illustration purposes, we assume that $\left( x_{\mathrm{p}}, y_{\mathrm{p}}, z_{\mathrm{p}}\right) = \left( 0.6, 0.4, 1.5 \right)$. The deposition velocities are set to be within the range of values reported in \cite{gao2007modeling} as follows: $d_{x_1}=10^{-8}\,  \mathrm{m/s^2}$, $d_{x_2}=10^{-5},  \mathrm{m/s^2}$, $d_{y_1}=10^{-4},  \mathrm{m/s^2}$, $d_{y_2}=10^{-6},  \mathrm{m/s^2}$, $d_{z_1}=10^{-1},  \mathrm{m/s^2}$, and $d_{z_2}=10^{-7},  \mathrm{m/s^2}$. We then evaluate the normalized viral concentration,  $\mathcal{C}/Q$, after 10 minutes, i.e., after the infected person left by 9 minutes. Fig. \ref{fig:bounds_x} plots $\mathcal{C}/Q$ versus the $x$-direction at $y=0.4\, \mathrm{m}$ and $z=1.5\, \mathrm{m}$. The performance versus the $y$-direction is then shown in Fig. \ref{fig:bounds_y} at $x=0.6\, \mathrm{m}$ and $z=1.5\, \mathrm{m}$. Finally, we plot $\mathcal{C}/Q$  performance versus the $z$-direction at $x=0.6\, \mathrm{m}$ and $y=0.4\, \mathrm{m}$ in Fig. \ref{fig:bounds_z}. In addition, to best characterize the spatial concentration performance, we compare the exact integral evaluation based on the circular planar modeling in \eqref{C_c1} to the integration bounds that are based on the square planar source model in \eqref{Css}. More specifically, we evaluate \eqref{Css} at three different squares: the large square of side length $2r_{\mathrm{c}}$ which gives an upper bound, the small square with side length $\sqrt{2}r_{\mathrm{c}}$ which gives a lower bound, and the square of side length $\sqrt{\pi}r_{\mathrm{c}}$ which gives an approximate concentration performance. Fig. \ref{fig:bounds_x}, Fig. \ref{fig:bounds_y}, and Fig. \ref{fig:bounds_z} visualize the upper and lower normalized concentration bounds, together with the exact and approximate values. The figures particularly highlight how tight is the approximate solution proposed in \eqref{Css}, as it gives a close performance to the exact solution \eqref{C_c}, despite the reduced computation burden of \eqref{Css}.

\begin{figure}[H]
	\centering
	\includegraphics[width=4.5in]{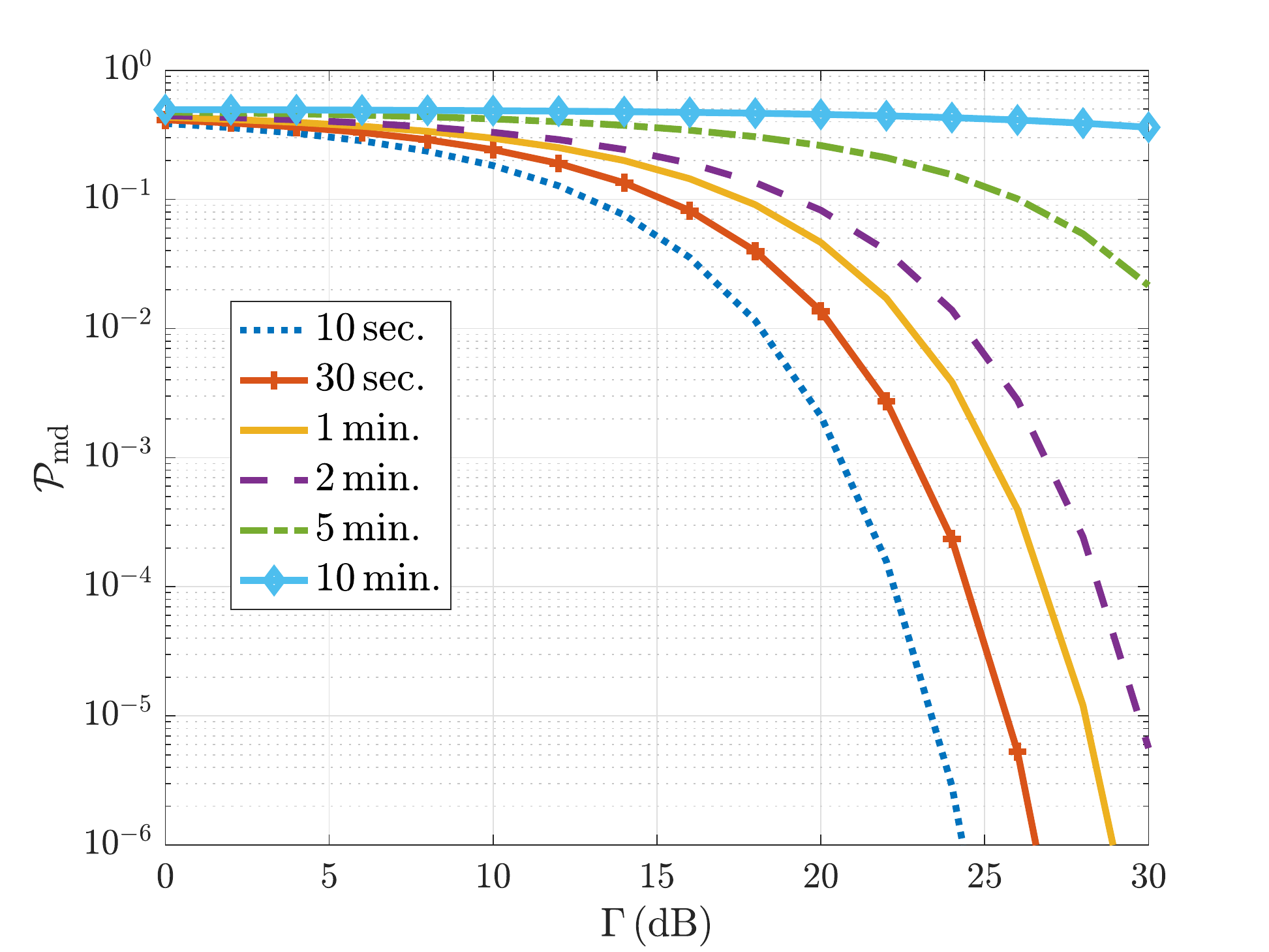}
	\caption{Probability of miss-detection versus $\Gamma$ at different time instances.} \label{fig:Pmd_Ex1}
\end{figure}

Fourthly, after evaluating the spatial-temporal concentration, we next explore the possibility of viral detection using electrical detectors. In this example, we assume the same environment setup used in the previous example. Further, to mimic the scenario of a short stay in an elevator-like environment, we assume that the infected person stays for 5 seconds. We then study the probability of miss-detection versus $\Gamma$ at different detection time instances, as illustrated in Fig. \ref{fig:Pmd_Ex1}. We assume the detector is located at $\left( 0.8, 0.1, 1.4 \right)$ and has $T_{\mathrm{s}} = 0.5\, \mathrm{sec.}$. We also set the side of the cubic sampling volume to $5\, \mathrm{cm}$. According to the results shown in Fig. \ref{fig:Pmd_Ex1}, viruses can be best detected if the detection time instance is close enough to the human existence in elevators. This is particularly possible in cases where detectors are supported with powerful abilities such as sampling efficiency, binding probability, and available viral concentration. For example, to detect viruses with $\mathcal{P}_{\mathrm{md}}$ of less than $10^{-2}$, we need to have detectors with  $\Gamma = 24\, \mathrm{dB}$. Such detectors would be able to detect the virus within 2 minutes, given that the infected person stays for a short time, as illustrated in Fig. \ref{fig:Pmd_Ex1}.

\begin{figure}
	\centering
	\includegraphics[width=4.5in]{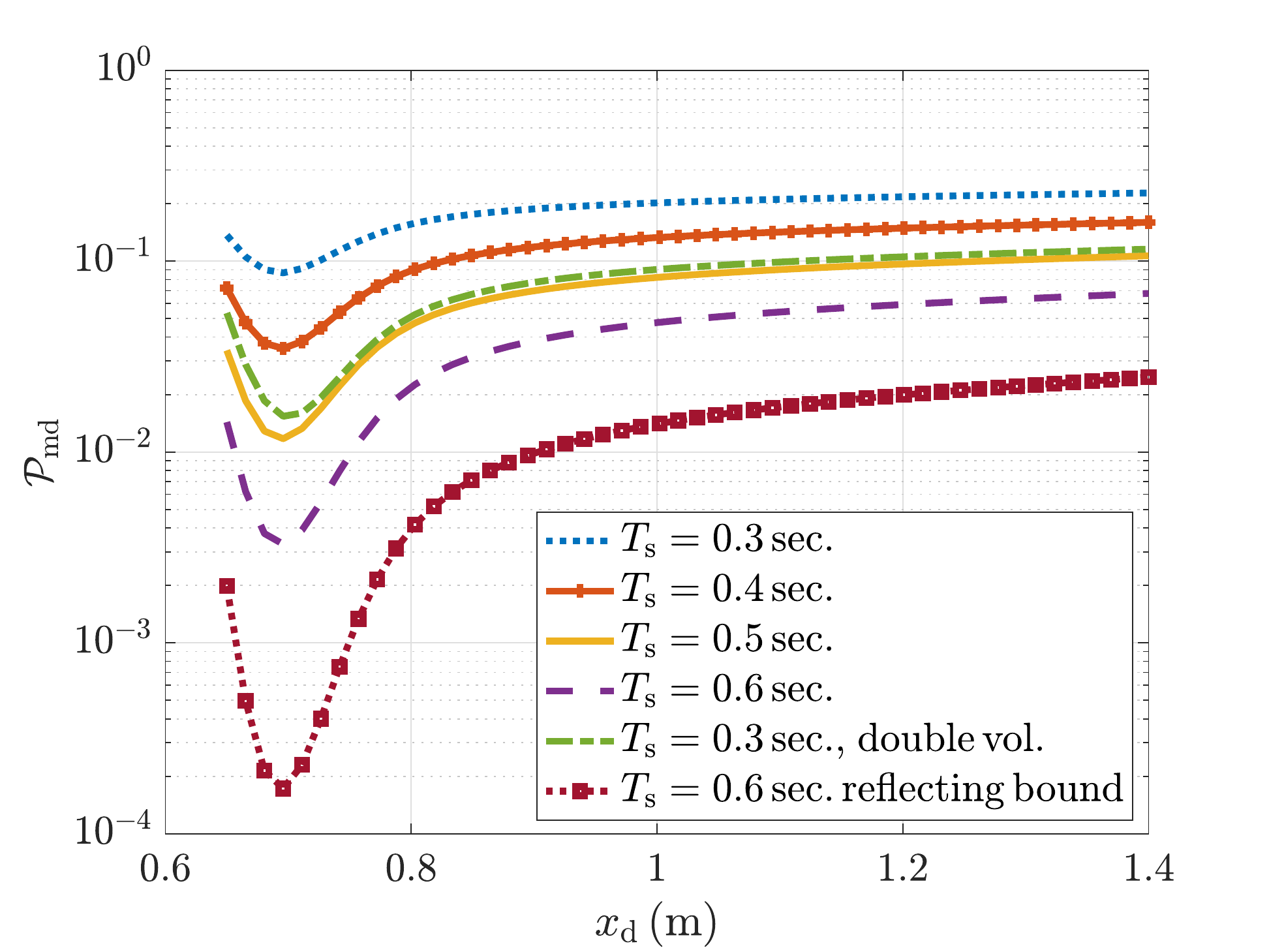}
	\caption{Probability of miss-detection versus detector location along $x$-direction.} \label{fig:Pmd_Ex2}
\end{figure}

Finally, we study the effect of detector location, sampling time, sampling volume, and reflecting boundary on the miss-detection probability. We use the same system setup parameters as in Fig. \ref{fig:Pmd_Ex1}, and evaluate in Fig. \ref{fig:Pmd_Ex2} $\mathcal{P}_{\mathrm{md}}$  versus the detector center location in the $x-$ direction after 1 minute for $\Gamma = 20 \, \mathrm{dB}$, different $T_{\mathrm{s}}$, double sampling volume, and a highly reflecting bound scenario. Fig. \ref{fig:Pmd_Ex2} shows that the detection process is efficient when the detector is close to the human location due to the slow diffusion process of the aerosol particles and the non-existence of an airflow. The figure also shows that slightly increasing the sampling time improves the performance significantly and provides robust detection abilities, especially for nearby locations. On the other hand, while increasing  $V_{\mathrm{s}}$ can improve the miss-detection performance and can allow using less $T_{\mathrm{s}}$, the performance gain due to doubling $T_{\mathrm{s}}$ is better than doubling $V_{\mathrm{s}}$, since increasing $T_{\mathrm{s}}$ allows collecting extra diffused viral particles. Another critical aspect illustrated in Fig. \ref{fig:Pmd_Ex2} is the impact of the reflecting/absorbing characteristics of the room boundaries. The figure shows that when the detector is close to a good reflecting boundary with $d_{y_1}=10^{-9}\, \mathrm{m/s^2}$, the viral aerosol particles can accumulate near the y-boundary, which allows the detector to collect extra samples, thereby improving the detection capability. For instance, having such reflecting abilities can provide a robust detection of $2\times10^{-4}$ at $x=0.7\, \mathrm{m}$, and a good performance up to $10^{-2}$ up to $x=0.9\, \mathrm{m}$, which illustrates the improved detection capabilities of the reflecting boundaries as compared to the partial absorption boundary case, as shown in Fig. \ref{fig:Pmd_Ex2}.

\section{Conclusion}

In light of the hazardous airborne transmission of pathogens, characterizing the viral concentration and modeling the viral transmission are considered significant milestones for enhancing viral detection capabilities. In this paper, we characterized the viral spatial-temporal concentration in a small bounded environment with partial reflecting/absorption boundaries. We derived a close form expression of the viral concentration using detailed analysis based on the mass conservation model and Fick's law. Moreover, we adopted an accurate model for the exhaled breath, which allowed us to model it as a continuous planner source. Although such an accurate model complicated the analysis, we managed to propose a simplified approximation for the viral concentration expression, which reduces the computation burden significantly and achieves a very close performance to the exact one. We then analyzed the abilities of electronic biosensors to detect the viruses from aerosol by analyzing the miss-detection probability. Finally, we presented several simulation scenarios that helped us understanding the airborne virus spread in a bounded environment and its associated miss-detection probability. We figured that the reflecting characteristics of the boundaries play a significant role in extending the existence of viral aerosol particles suspended in the room, which is further validated through improving the detection ability.

\appendices
\section{Proof of equivalent PDE in the $\nu-\mathrm{th}$ dimension} \label{Appendix_A}
First, we consider the main PDE in \eqref{main_pde1} and evaluate it using \eqref{C_sep} and obtain
\begin{equation} \label{App_A}
\begin{split}
&\frac{\partial  \mathcal{C}_{x} \! \left(x,  t\right)}{\partial t}  \mathcal{C}_{y} \! \left(y,  t\right) \mathcal{C}_{z} \! \left( z, t\right) +\mathcal{C}_{x} \! \left(x,  t\right)  \frac{\partial \mathcal{C}_{y}\! \left(y,  t\right)}{\partial t}  \mathcal{C}_{z} \!\left( z, t\right) \\ & +  \mathcal{C}_{x} \! \left( x, t\right)  \mathcal{C}_{y} \! \left(y,  t\right)\frac{\partial \mathcal{C}_{z}\! \left(z,  t\right)}{\partial t}  =  K_{x} \frac{\partial^2 \mathcal{C}_{x}}{\partial x^2} \!\left(x,  t\right)  \mathcal{C}_{y} \! \left(y,  t\right) \mathcal{C}_{z} \! \left( z, t\right)  \\&+ K_{y} \mathcal{C}_{x}\!\left(x,  t\right)  \frac{\partial^2 \mathcal{C}_{y} \! \left(y,  t\right) }{\partial y^2} \mathcal{C}_{z} \! \left(z,  t\right)  \\&+ K_{z}\mathcal{C}_{x}\!\left(x,  t\right) \mathcal{C}_{x}\!\left(y,  t\right)  \frac{\partial^2 \mathcal{C}_{z}\!\left(z,  t\right)}{\partial z^2}.
\end{split}
\end{equation}
Then, we divide both sides of \eqref{App_A} by $\mathcal{C}_{\mathrm{p}} \left( x, y, z, t \right)$ using the proposed form in \eqref{C_sep} obtaining
\begin{equation} \label{App_A1}
\begin{split}
&\frac{1}{\mathcal{C}_{x} \! \left(x,  t\right)} \frac{\partial  \mathcal{C}_{x} \! \left(x,  t\right)}{\partial t}  + \frac{1}{\mathcal{C}_{y} \! \left(y,  t\right)} \frac{\partial  \mathcal{C}_{y} \! \left(y,  t\right)}{\partial t}  + \frac{1}{\mathcal{C}_{z} \! \left(z,  t\right)} \frac{\partial  \mathcal{C}_{z} \! \left(z,  t\right)}{\partial t}  \\
&=  \frac{K_x}{\mathcal{C}_{x} \! \left(x,  t\right)} \frac{\partial^2 \mathcal{C}_{x} \!\left(x,  t\right) }{\partial x^2} +  \frac{K_y}{\mathcal{C}_{y} \! \left(y,  t\right)} \frac{\partial^2 \mathcal{C}_{y} \!\left(y,  t\right) }{\partial y^2}  \\
& +  \frac{K_z}{\mathcal{C}_{z} \! \left(z,  t\right)} \frac{\partial^2 \mathcal{C}_{z} \!\left(z,  t\right) }{\partial z^2} .
\end{split}
\end{equation}
After comparing both sides of \eqref{App_A1}, we find
\begin{equation} \label{App_A2}
\begin{split}
\frac{1}{\mathcal{C}_{x} \! \left(x,  t\right)} \frac{\partial  \mathcal{C}_{x} \! \left(x,  t\right)}{\partial t}  & =  \frac{K_x}{\mathcal{C}_{x} \! \left(x,  t\right)} \frac{\partial^2 \mathcal{C}_{x} \!\left(x,  t\right) }{\partial x^2} \\
\frac{1}{\mathcal{C}_{y} \! \left(y,  t\right)} \frac{\partial  \mathcal{C}_{y} \! \left(y,  t\right)}{\partial t}  & =  \frac{K_y}{\mathcal{C}_{y} \! \left(y,  t\right)} \frac{\partial^2 \mathcal{C}_{y} \!\left(y,  t\right) }{\partial y^2} \\
\frac{1}{\mathcal{C}_{z} \! \left(z,  t\right)} \frac{\partial  \mathcal{C}_{z} \! \left(z,  t\right)}{\partial t}  & =  \frac{K_z}{\mathcal{C}_{z} \! \left(z,  t\right)} \frac{\partial^2 \mathcal{C}_{z} \!\left(z,  t\right) }{\partial z^2},
\end{split}
\end{equation}
which is written in a general form in \eqref{sub_PDE}.

\section{Positive Eigenvector Weight Derivation} \label{Appendix_B}
First, we evaluate the numerator integration of \eqref{weight} from,
\begin{align}
{\int_0^{L_\nu} \!\!  \mathcal{C}_{\nu} \left(  \nu, t_0  \right)   \Phi_n \left(  \nu, t_0  \right) d \nu} = & \int_0^{L_\nu}  \!\!   Q_\nu   \delta(\nu- \nu_{\mathrm{p}} )   \Phi_n \left(  \nu, t_0  \right) d \nu  \nonumber  \\
=  & \; \,   \Phi_n  \left(   \nu_{\mathrm{p}}, t_0  \right)   \nonumber  \\
\!\!= Q_\nu    \Big(  \! \cos(\lambda_{\nu,n} \nu_{\mathrm{p}})     +   \frac{\beta_{\nu_1}}{\lambda_{\nu,n}} &  \sin(\lambda_{\nu,n}\nu_{\mathrm{p}})\! \Big)   e^{-K_\nu \lambda^2_{\nu,n} t_0}. \label{App_B_num}
\end{align}
Then, we find the denominator integration of \eqref{weight} from
\begin{align}
& \int_0^{L_\nu}  \Phi^2_n \left(  \nu, t_0  \right)   d \nu \nonumber \\
& = \int_0^{L_\nu} \!\!\!  \Big(  \! \cos(\lambda_{\nu,n} \nu)     +   \frac{\beta_{\nu_1}}{\lambda_{\nu,n}}   \sin(\lambda_{\nu,n}\nu)\! \Big)^{2}  e^{-2K_\nu \lambda^2_{\nu,n} t_0}   d \nu \nonumber \\
& =  e^{-2K_\nu \lambda^2_{\nu,n} t_0} \bigg[  \int_0^{L_\nu}  \left(  \cos^{2}(\lambda_{\nu,n} \nu)    +   \frac{\beta^2_{\nu_1}}{\lambda^2_{\nu,n}}   \sin^2(\lambda_{\nu,n}\nu) \right)  d \nu  \nonumber \\
& \qquad + \frac{2\beta_{\nu_1}}{\lambda_{\nu,n}}  \int_0^{L_\nu}  \cos(\lambda_{\nu,n} \nu)   \sin(\lambda_{\nu,n}\nu) \, d \nu \bigg].
\end{align}
Then, we use the trigonometric identities, $\cos(2\theta)=2\cos^2(\theta) - 1 = 1 - 2\sin^2(\theta)$  and $\sin(2\theta)=2\sin(\theta)\cos(\theta)$, we can write the integration equivalently  as
\begin{align}
& \int_0^{L_\nu} \!\!\!\! \Phi^2_n \left(  \nu, t_0  \right)   d \nu =  e^{-2K_\nu \lambda^2_{\nu,n} t_0} \bigg[  \frac{1}{2}  \int_0^{L_\nu} \!\!\!\! \left( 1+  \cos(2\lambda_{\nu,n} \nu)  \right)  d\nu  \nonumber \\
&   +   \frac{\beta^2_{\nu_1}}{2\lambda^2_{\nu,n}}    \int_0^{L_\nu} \!\!\!\!\!\! \left( 1 -  \cos(2\lambda_{\nu,n} \nu)  \right)  d\nu   + \frac{\beta_{\nu_1}}{\lambda_{\nu,n}}  \int_0^{L_\nu}  \!\!\!\!\!\!     \sin(2\lambda_{\nu,n}\nu) \, d \nu \bigg]  \nonumber \\
&=  e^{-2K_\nu \lambda^2_{\nu,n} t_0} \bigg[  \frac{L_\nu}{2}  + \frac{\sin(2\lambda_{\nu,n} L_\nu) }{4\lambda_{\nu, n}}    + \frac{L_\nu \beta^2_{\nu_1}}{2\lambda^2_{\nu,n}}  \nonumber \\
& - \frac{\beta^2_{\nu_1} \sin(2\lambda_{\nu,n} L_\nu) }{4\lambda^3_{\nu, n}}   + \frac{\beta_{\nu_1}}{2\lambda^2_{\nu,n}} \left( 1 - \cos(2\lambda_{\nu,n} L_\nu) \right) \bigg] \nonumber \\
& =  \frac{e^{-2K_\nu \lambda^2_{\nu,n} t_0}}{4 \lambda^3_{\nu, n}} \bigg[ 2L_\nu \lambda^3_{\nu, n} +  \lambda^2_{\nu, n}\sin(2\lambda_{\nu,n} L_\nu) +   2L_\nu \lambda_{\nu, n} \beta^2_{\nu_1}  \nonumber \\
& - \beta^2_{\nu_1} \sin(2\lambda_{\nu,n} L_\nu) + {2\lambda_{\nu, n} \beta_{\nu_1}} \left( 1 - \cos(2\lambda_{\nu,n} L_\nu) \right) \bigg] \nonumber \\
& =  \frac{e^{-2K_\nu \lambda^2_{\nu,n} t_0}}{4 \lambda^3_{\nu, n}} \bigg[  \left( \lambda^2_{\nu, n} - \beta^2_{\nu_1} \right) \sin(2\lambda_{\nu,n} L_\nu)  \nonumber  \\
&- {2\lambda_{\nu, n} \beta_{\nu_1}}   \cos(2\lambda_{\nu,n} L_\nu)    + 2\lambda_{\nu, n} \left(  \left(\lambda_{\nu, n}^2+\beta_{\nu_1}^2\right) L_\nu +\beta_{\nu_1}  \right)  \bigg].  \label{App_B_den}
\end{align}
After evaluating the weights numerator and denominator integration in \eqref{App_B_num} and \eqref{App_B_den}, respectively, to obtain \eqref{weight}.

\section{Negative Eigenvector Weight Derivation} \label{Appendix_B_n}
Similar to the positive eigenvalue case, we evaluate the numerator integration of \eqref{weight} from,
\begin{align}
{\int_0^{L_\nu} \!\!  \mathcal{C}_{\nu} \left(  \nu, t_0  \right)   \tilde{\Phi}   \left(  \nu, t_0  \right) d \nu} = & \int_0^{L_\nu}  \!\!   Q_\nu   \delta(\nu- \nu_{\mathrm{p}} )   \tilde{\Phi}  \left(  \nu, t_0  \right) d \nu  \nonumber  \\
=  & \; \,   \tilde{\Phi}  \left(   \nu_{\mathrm{p}}, t_0  \right)   \nonumber  \\
\!\!= \Big(  \cosh(\tilde{\lambda}_\nu \nu_{\mathrm{p}}) +\dfrac{\beta_{\nu_1}}{ {\tilde{\lambda}}_{\nu}} & \sinh(\tilde{\lambda}_\nu \nu_{\mathrm{p}}) \Big) e^{-K_\nu\tilde{\lambda}^2_{\nu} t}. \label{App_B_num_n}
\end{align}
Then, we find the denominator integration of \eqref{weight} from
\begin{align}
& \int_0^{L_\nu}  \tilde{\Phi} \left(  \nu, t_0  \right)   d \nu \nonumber \\
& = \int_0^{L_\nu} \!\!\!  \Big(  \! \cosh( \tilde{\lambda}_{\nu} \nu)     +   \frac{\beta_{\nu_1}}{\tilde{\lambda}_{\nu}}   \sinh(\tilde{\lambda}_{\nu}\nu)\! \Big)^{2}  e^{-2K_\nu \tilde{\lambda}^2_{\nu} t_0}   d \nu \nonumber \\
& =  e^{-2K_\nu \tilde{\lambda}^2_{\nu} t_0} \bigg[  \int_0^{L_\nu}  \left(  \cosh^{2}(\tilde{\lambda}_{\nu} \nu)    +   \frac{\beta^2_{\nu_1}}{\tilde{\lambda}^2_{\nu}}   \sinh^2(\tilde{\lambda}_{\nu}\nu) \right)  d \nu  \nonumber \\
& \qquad + \frac{2\beta_{\nu_1}}{ \tilde{\lambda}_{\nu}}  \int_0^{L_\nu}  \cosh(\tilde{\lambda}_{\nu} \nu)   \sinh(\tilde{\lambda}_{\nu}\nu) \, d \nu \bigg].
\end{align}
Then, we use the trigonometric identities, $\cosh (2\theta)=2\cosh^2(\theta) - 1 = 2\sinh^2(\theta)+1 $  and $\sinh(2\theta)=2\sinh(\theta)\cosh(\theta)$, we can write the integration equivalently~as
\begin{align}
& \int_0^{L_\nu} \!\!\!\! \tilde{\Phi}^2 \left(  \nu, t_0  \right)   d \nu =  e^{-2K_\nu \tilde{\lambda}^2_{\nu} t_0} \bigg[  \frac{1}{2}  \int_0^{L_\nu} \!\!\!\! \left( 1+  \cosh(2\tilde{\lambda}_{\nu} \nu)  \right)  d\nu   +   \frac{\beta^2_{\nu_1}}{2\tilde{\lambda}^2_{\nu}}    \int_0^{L_\nu} \!\!\!\!  \left(   \cosh(2\tilde{\lambda}_{\nu} \nu) - 1  \right)  d\nu  \nonumber \\ & \qquad \qquad \qquad \qquad \qquad \qquad \qquad \qquad \qquad \qquad \qquad \qquad \qquad + \frac{\beta_{\nu_1}}{\lambda_{\nu,n}}  \int_0^{L_\nu}  \!\!\!\!\!\!     \sinh(2 \tilde{\lambda}_{\nu}\nu) \, d \nu \bigg]  \nonumber \\
&=  e^{-2K_\nu \tilde{\lambda}^2_{\nu} t_0} \bigg[  \frac{L_\nu}{2}  + \frac{\sinh(2\tilde{\lambda}_{\nu} L_\nu) }{4\tilde{\lambda}_{\nu}}    -  \frac{L_\nu \beta^2_{\nu_1}}{2\tilde{\lambda}^2_{\nu}}  + \frac{\beta^2_{\nu_1} \sinh(2\tilde{\lambda}_{\nu} L_\nu) }{4\tilde{\lambda}^3_{\nu}}   + \frac{\beta_{\nu_1}}{2\tilde{\lambda}^2_{\nu}} \left(  \cosh(2 \tilde{\lambda}_{\nu} L_\nu) - 1  \right) \bigg] \nonumber \\
& =  \frac{e^{-2K_\nu \tilde{\lambda}^2_{\nu} t_0}}{4 \tilde{\lambda}^3_{\nu}} \bigg[ 2L_\nu \tilde{\lambda}^3_{\nu} +  \tilde{\lambda}^2_{\nu} \sinh(2 \tilde{\lambda}_{\nu} L_\nu)  - 2L_\nu \tilde{\lambda}_{\nu} \beta^2_{\nu_1}    + \beta^2_{\nu_1} \sinh(2 \tilde{\lambda}_{\nu} L_\nu) + {2 \tilde{\lambda}_{\nu} \beta_{\nu_1}} \left(   \cosh(2 \tilde{\lambda}_{\nu} L_\nu) - 1  \right)  \bigg] \nonumber \\
& =  \frac{e^{-2K_\nu \lambda^2_{\nu,n} t_0}}{4 \lambda^3_{\nu}} \bigg[  \left( \tilde{\lambda}^2_{\nu} + \beta^2_{\nu_1} \right) \sinh(2\tilde{\lambda}_{\nu} L_\nu)  \nonumber  \\
&+ {2\tilde{\lambda}_{\nu} \beta_{\nu_1}}   \cosh(2\tilde{\lambda}_{\nu} L_\nu)    + 2\tilde{\lambda}_{\nu} \left(  \left(\tilde{\lambda}_{\nu}^2 - \beta_{\nu_1}^2\right) L_\nu  - \beta_{\nu_1}  \right)  \bigg].  \label{App_B_den_n}
\end{align}
Thus, we can find the weights numerator and denominator integration in \eqref{App_B_num_n} and \eqref{App_B_den_n}, respectively, to obtain \eqref{weight_n}.

\end{document}